\documentclass{aa}  

\usepackage{graphicx}
\usepackage{txfonts}
\usepackage{mathpazo}
\usepackage[T1]{fontenc}
\usepackage{xcolor}
\usepackage{appendix}
\usepackage{multicol}
\usepackage[flushleft]{threeparttable}
\usepackage{array,booktabs}
\usepackage{amsmath}
\usepackage{textcomp}
\usepackage{caption}
\usepackage{subcaption}
\usepackage{hyperref}
\hypersetup{
    colorlinks=true,
    linkcolor=blue,
    filecolor=red,
    citecolor=blue,
    urlcolor=blue,
}

\begin{document} 

   \title{Linking ice and gas in the Coronet cluster in Corona Australis}

    \author{G. Perotti
          \inst{1} \fnmsep \inst{2},
          J. K. J{\o}rgensen\inst{1},
          W. R. M. Rocha\inst{3},
          A. Plunkett\inst{4},
          E. Artur de la Villarmois\inst{5} \fnmsep \inst{6},
          L. E. Kristensen\inst{1},
          M. Sewi{\l}o\inst{7} \fnmsep \inst{8} \fnmsep \inst{9},
          P. Bjerkeli\inst{10},
          H. J. Fraser\inst{11}, and
          S. B. Charnley\inst{12}
        }
    \institute{Niels Bohr Institute, University of Copenhagen, {\O}ster Voldgade 5$-$7, 1350 Copenhagen K, Denmark
    \and Max Planck Institute for Astronomy, K{\"o}nigstuhl 17, D-69117 Heidelberg, Germany \\
              \email{perotti@mpia.de}
              \and Laboratory for Astrophysics, Leiden Observatory, Leiden University, PO Box 9513, 2300 RA Leiden, The Netherlands
              \and National Radio Astronomy Observatory, 520 Edgemont Rd, Charlottesville, VA 22903-2475, USA
              \and Instituto de Astrof{\'i}sica, Ponticia Universidad Cat{\'o}lica de Chile, Av. Vicu{\~{n}}a Mackenna 4860, 7820436 Macul, Santiago, Chile
              \and N{\'u}cleo Milenio de Formaci{\'o}n Planetaria (NPF), Chile
               \and Exoplanets and Stellar Astrophysics Laboratory, NASA Goddard Space Flight Center, Greenbelt, MD 20771, USA
              \and Department of Astronomy, University of Maryland, College Park, MD 20742, USA
              \and Center for Research and Exploration in Space Science and Technology, NASA Goddard Space Flight Center, Greenbelt, MD 20771, USA
              \and Department of Space, Earth, and Environment, Chalmers University of Technology, Onsala Space Observatory, 439 92 Onsala, Sweden
              \and School of Physical Sciences, The Open University, Walton Hall, Milton Keynes, MK7 6AA, United Kingdom
              \and Astrochemistry Laboratory, NASA Goddard Space Flight Center, Greenbelt, MD 20771, USA
       }
   \date{Received ; accepted}
 
\abstract
  {During the journey from the cloud to the disc, the chemical composition of the protostellar envelope material can be either preserved or processed to varying degrees depending on the surrounding physical environment.}
  {This works aims to constrain the interplay of solid (ice) and gaseous methanol (CH$_3$OH) in the outer regions of protostellar envelopes located in the Coronet cluster in Corona Australis (CrA), and assess the importance of irradiation by the Herbig Ae/Be star R~CrA. CH$_3$OH is a prime test-case as it predominantly forms as a consequence of the solid-gas interplay (hydrogenation of condensed CO molecules onto the grain surfaces) and it
  plays an important role in future complex molecular processing.}
   {We present 1.3~mm Submillimeter Array (SMA) and Atacama Pathfinder Experiment (APEX) observations towards the envelopes of four low-mass protostars in the Coronet. Eighteen molecular transitions of seven species are identified. We calculate CH$_3$OH gas-to-ice ratios in this strongly irradiated cluster and compare them with ratios determined towards protostars located in less irradiated regions such as the Serpens SVS~4 cluster in Serpens Main and the Barnard 35A cloud in the $\lambda$ Orionis region.}
   {The CH$_3$OH gas-to-ice ratios in the Coronet vary by one order of magnitude (from 1.2$\times$10$^{-4}$ to 3.1$\times$10$^{-3}$) which is similar to less irradiated regions as found in previous studies. We find that the CH$_3$OH gas-to-ice ratios estimated in these three regions are remarkably similar despite the different UV radiation field intensities and formation histories.}
   {This result suggests that the overall CH$_3$OH chemistry in the outer regions of low-mass envelopes is relatively independent of variations in the physical conditions and hence that it is set during the prestellar stage.}

   \keywords{ISM: molecules --- stars:protostars --- astrochemistry --- molecular processes --- ISM:individual objects: R~CrA}
   
    \titlerunning{Linking ice and gas toward the Coronet cluster in Corona Australis}
    \authorrunning{G. Perotti et al.}
  
   \maketitle
   
%

\section{Introduction}
Solar-type stars and planets form inside molecular clouds, when dense cores undergo gravitational collapse. While collapsing, the gas and dust constituting the clouds is assembled into infalling envelopes, streamers and circumstellar discs, which supply the fundamental ingredients for planet formation (see e.g., \citealt{Pineda2023}). Recent observations (e.g., \citealt{Andrews2018,Harsono2018,Keppler2018,Segura-Cox2020}) suggest that planets form earlier than previously thought (a few 10$^5~$years; \citealt{Tychoniec2020}) and in tandem with their host stars \citep{Alves2020}. In this context, it is still unclear whether the composition of the molecular cloud material is preserved when becoming part of planets or, instead, entirely thermally processed losing the natal cloud's chemical fingerprint (see e.g., \citealt{Jorgensen2020,vanDishoeck2020} for recent reviews). 

To address this query it is necessary to study how the environment impacts the physical and chemical properties of embedded protostars and their discs. For instance, external irradiation \citep{Winter2020} and cosmic ray ionization \citep{Kuffmeier2020} might shape the chemical and physical evolution of forming low-mass stars by e.g., photoevaporating their discs (\citealt{vanTerwisga2020,Haworth2021} and references therein) or leading to less massive discs \citep{Cazzoletti2019}. In this paper we investigate the variations of the methanol gas and ice towards deeply embedded sources in Corona Australis, to study the effects of external irradiation on the gas-to-ice ratios.

CH$_3$OH is the most suitable molecule for the main aim of this work because it is abundantly detected in both solid and gas phases \citep{Boogert2015,McGuire2022}. Most importantly, it is primarily formed in the solid-state \citep{Watanabe2002,Fuchs2009,Qasim2018,Simons2020,Santos2022} as its gas-phase formation pathways are considerably less efficient \citep{Roberts2000,Garrod2006,Geppert2006}. Finally, CH$_3$OH is regarded as the gateway species for the formation of complex organic molecules both in the solid-state (e.g., \citealt{Oberg2009b,Chuang2016,Fedoseev2017}) and in the gas phase (e.g., \citealt{Shannon2014,Balucani2015}).

\begin{figure}
\centering
\includegraphics[width=\hsize]{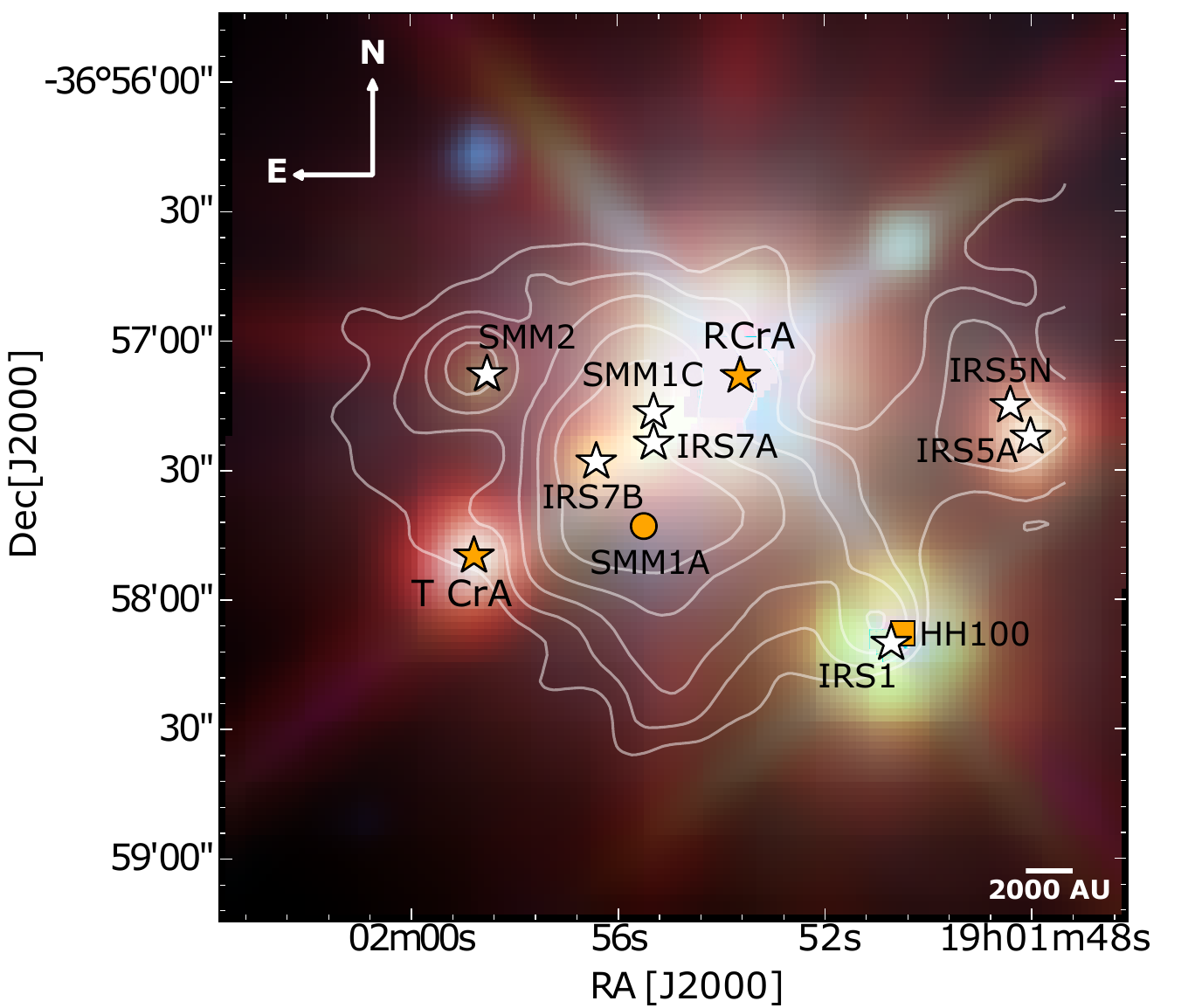}
\caption{Three-color image of the Coronet cluster (\textit{WISE} 3.4~$\mu$m (blue), 4.6~$\mu$m (green), and 12~$\mu$m (red) bands; \citealt{Wright2010}) overlaid with SCUBA 850~$\mu$m density flux \citet{Nutter2005}; contours are in decreasing steps starting at the peak flux (3.7~Jy beam$^{-1}$) and subtracting 30\% from the previous level. The white stars mark the positions of Class 0/I YSOs in the R~CrA/Coronet region \citep{Peterson2011,Nutter2005}, whereas the orange stars indicate the Herbig Ae/Be star R~CrA and the T Tauri star T~CrA. The pre-stellar core candidate SMM1A is indicated with an orange circle and the orange square represents the Herbig Haro object HH100.}
\label{rgb_rcra}
\end{figure}

Corona Australis (CrA) is one of the nearest regions with ongoing star-formation, located at a distance of 149.4$\pm$0.4~pc as estimated from \textit{Gaia}-DR2 measurements \citep{Galli2020}. The most recent census of the cloud counts 393 young stellar object (YSO) candidates \citep{Esplin2022}: these are relatively evolved, mainly Class II and III sources, and they are predominantly concentrated in the most extincted region, the "head" \citep{Peterson2011,Alves2014}. The correlation between the star formation activity and the head-tail structure of Corona Australis has been investigated by \citet{Sami2019}, who found that the spatial distribution of dense cores is a consequence of the physical conditions of the large scale environment present at the time the cloud assembled. The youngest population of YSOs (Class 0/I) is situated in the Coronet cluster \citep{Taylor1984,Forbrich2007a,Forbrich2007b}, also associated with the luminous Herbig Ae/Be star R~CrA \citep{Peterson2011}. 

Figure~\ref{rgb_rcra} displays a section of the Coronet cluster and a summary of the principal cluster members is provided in Table~\ref{table:samples_sources}. The Herbig Ae/Be R~CrA (spectral type B5$-$B8; \citealt{Gray2006,Bibo1992}) is the brightest star in this very young cluster that has an estimated age of 0.5$-$1~Myr \citep{Sicilia-Aguilar2011}. Due to the variable nature of R~CrA, its stellar mass and luminosity are uncertain \citep{Mesa2019}. Recent \textit{Very Large Telescope} (VLT)/SPHERE observations of R~CrA resolved a companion at a separation of $\sim$0.156$''-$0.184$''$ and complex extended jet-like structures around the star \citep{Mesa2019}. 
A second variable star, the T Tauri star T~CrA is present in the Coronet cluster, approximately 30$\arcsec$ to the south-east of R~CrA \citep{Herbig1960,Taylor1984}.
The region surrounding the two variable stars harbours seven identified Class 0/I YSOs (Table~\ref{table:samples_sources}). 

The Coronet cluster members have been at the center of active multi-wavelength research, mostly aimed at the characterization of the properties of YSOs at very early stages of stellar evolution. Some of the most studied objects are IRS7B, IRS7A, SMM1A, SMM1C and SMM2 \citep{Nutter2005,Groppi2007,Miettinen2008,Chen2010, Peterson2011}. The spectral energy distributions (SEDs) of the cluster members have been investigated by \citet{Groppi2007}, suggesting that SMM1C is a Class~0 YSO and IRS7B is a transitional Class~0/I object. IRS7A and SMM2 are likely Class I sources \citep{Peterson2011} and SMM1A is classified as a pre-stellar core candidate \citep{Nutter2005,Chen2010}. IRS1, IRS5N and IRS5A are suggested as Class~I YSO based on submillimeter and infrared observations \citep{Peterson2011}. 

\begin{table*}
\begin{center}
\caption{Overview of the Coronet cluster objects.}
\label{table:samples_sources}
\renewcommand{\arraystretch}{1.4}
\begin{tabular}{lccccc} 
\hline \hline
Object  &  RA       &  DEC          &    Designation$^{(a)}$     & Other names$^{(b)}$ \\  
        & [J2000]   & [J2000]       &                     &            \\ \hline
IRS5A   & 19:01:48.03 & -36:57:22.2 &    Class I YSO   & CrA-19 \\
IRS5N   & 19:01:48.46 & -36:57:14.7 &    Class I YSO   & CrA-20 \\
HH100~IRS1  & 19:01:50.56 & -36:58:08.9& Herbig-Haro object   & $-$\\
IRS1    & 19:01:50.68 & -36:58:09.7 &    Class I YSO          & VSSt~15, TS~2.6\\
R~CrA   & 19:01:53.67 & -36:57:08.0 &    Herbig Ae/Be star    & $-$ \\
SMM1C   & 19:01:55.29 & -36:57:17.0 &    Class I YSO & B9, Brown 9 \\ 
IRS7A   & 19:01:55.32 & -36:57:21.9 &    Class I YSO          & IRS7W, IRS7 \\ 
SMM1A   & 19:01:55.60 & -36:57:43.0 &    Pre-stellar core     & $-$      \\
IRS7B   & 19:01:56.40 & -36:57:28.3 &    Class I YSO & SMM1B, IRS7E \\
SMM2    & 19:01:58.54 & -36:57:08.5 &    Class I YSO & CrA-43, WMB55 \\
T~CrA   & 19:01:58.78 & -36:57:49.9 &    T Tauri star & $-$   \\
\hline
\end{tabular}
\end{center}
\footnotesize{\textbf{Notes.} All the coordinates are taken from \citet{Peterson2011}, except for SMM1A which comes from \citet{Nutter2005} and HH100~IRS1 from \citet{Boogert2008}. $^{(a)}$ The YSO class is assigned from the spectral index $\alpha$ reported in \citet{Peterson2011}; the objects designated as "Class I YSO" are Class I or younger. $^{(b)}$ From \citet{Peterson2011} and \citet{Lindberg2012}.}
\end{table*}

Along with the identification and the evolutionary stage designation of the YSOs in the Coronet, their chemical evolution has been the subject of a large number of studies over the last decades. The line-rich spectra of IRS7B and IRS7A have been investigated with the \textit{Atacama Pathfinder EXperiment} (APEX) single-dish telescope by \citet{Schoier2006}, who reported different kinetic temperatures for formaldehyde (H$_2$CO) and methanol (>30~K for H$_2$CO and $\approx$20~K for CH$_3$OH). \citet{Lindberg2012} suggested that the high temperatures in the region (>30~K) traced by the emission of H$_2$CO, are caused by external irradiation from the Herbig Ae/Be star R~CrA. With the purpose of analyzing the impact of external irradiation on the molecular inventory of low-mass protostars, systematic unbiased line surveys have been carried out firstly towards IRS7B with the \textit{Atacama Submillimeter Telescope Experiment} (ASTE) by \citet{Watanabe2012}, and extended to other Coronet cluster members using APEX by \citet{Lindberg2015}. These systematic surveys confirmed that external irradiation can affect the chemical nature of protostars, particularly by enhancing the abundances of Photon-Dominated Regions (PDRs) tracers (such as CN, C$_2$H, and c-C$_3$H$_2$). High-resolution \textit{Atacama Large Millimeter/submillimeter Array} (ALMA) observations found a low column density of CH$_3$OH in the inner hot regions of the protostellar envelopes which could either be due to the suppressed CH$_3$OH formation in ices at the higher temperatures in the region or due to the envelope structures on small scales being affected by the presence of a disc \citep{Lindberg2014b}.
The latter interpretation is supported by recent observations \citep{Artur_2018,vanGelder2022} and models \citep{Nazari2022a}. 

Along with the gas-phase molecular species, the composition of interstellar ices towards a few objects in the Coronet cluster (IRS5A, IRS5B, HH100~IRS1, IRS7A, IRS7B) has been modelled \citep{Siebenmoregen1997} and investigated in the near-infrared (3$-$5~$\mu$m) with the \textit{Very Large Telescope} \citep{Pontoppidan2003b}, in the mid-infrared (5$-$8~$\mu$m) using the \textit{Infrared Space Observatory} (\citealt{Keane2001}) and as part of the \textit{Spitzer} Space Telescope c2d Legacy Program (5$-$8~$\mu$m by \citealt{Boogert2008}; 8$-$10~$\mu$m by \citealt{Bottinelli2010}, and 14.5$-$16.5~$\mu$m by \citealt{Pontoppidan2008}). Ice features attributed to H$_2$O, CO, CO$_2$, NH$_3$, CH$_3$OH were securely identified in the astronomical spectra of the targeted sources. 

This paper explores the variations of CH$_3$OH gas and ice abundances, investigating the effects of external irradiation on the gas-phase and solid-state chemistries of the youngest objects in the Coronet cluster. We present \textit{Submillimeter Array} (SMA) observations towards five Coronet cluster members, thus complementing previous observations at 1.3~mm \citep{Lindberg2012} by increasing the number of detected chemical species in this region. To probe the CH$_3$OH large-scale emission, an APEX 150$'' \times$150$''$ map of the CH$_3$OH $J_K = 5_K-4_K$ Q-branch is also presented. The column densities of H$_2$O and CH$_3$OH ice are taken from \citet{Boogert2008}. Furthermore, the dependence of gas-to-ice ratios on the physical conditions are addressed, by determining gas-to-ice ratios of CH$_3$OH in this strongly irradiated cluster and comparing them with ratios obtained towards less irradiated star-forming regions \citep{Perotti2020,Perotti2021}. 

The paper is laid out as follows. Section~\ref{sec:obs} describes both the SMA and APEX observational setups, and the data reduction strategy. In Section~\ref{sec:results}, we present our key observational results, while the variations on the ice and gas abundance are analysed in Section~\ref{sec:analysis}. Section~\ref{sec:discussion} discusses the determined CH$_3$OH gas-to-ice ratios and Section~\ref{sec:conclusions} summarises our conclusions and lists avenues for future studies.

\begin{table*}
\begin{center}
\caption{Identified methanol transitions in the SMA and APEX data sets.}
\label{table:spectral_data_methanol}
\renewcommand{\arraystretch}{1.6}
\begin{tabular}{l l c c c c c }
\hline \hline
Species & Transition &  Frequency$^{(a)}$ & $A_\mathrm{ul}^{(a)}$  & $g_\mathrm{u}^{(a)}$ & $E_\mathrm{u}^{(a)}$ & $n_\mathrm{cr}^{(b)}$  \\ 
        &            & [GHz]            & [s$^{-1}$]&                               & [K]  & [cm$^{-3}$]\\ \hline 
CH$_3$OH  & 5$_{+0}~-~4_{+0}$  E  & 241.700     &  6.04~$\times~10^{-5}$  &11 & 47.9  &  1.7~$\times~10^6$ \\
          & 5$_{-1}~-~4_{-1}$  E  & 241.767     &  5.81~$\times~10^{-5}$  &11 & 40.4  & 4.8~$\times~10^7$\\
          & 5$_{0}~-~4_{0}$  A$^+$    & 241.791   &  6.05~$\times~10^{-5}$  &11 & 34.8  &4.7~$\times~10^5$\\
          & 5$_{+1}~-~4_{+1}$ E   & 241.879     &  5.96~$\times~10^{-5}$  &11 & 55.9  & 1.5~$\times~10^7$\\
          & 5$_{-2}~-~4_{-2}$  E  & 241.904     &  5.09~$\times~10^{-5}$  &11 & 60.7  & 2.3~$\times~10^8$ \\ \hline
\hline
\end{tabular}
\end{center}
\begin{center}
\begin{tablenotes}
\item{\textbf{Notes.}$^{(a)}$ The spectroscopic data are from \citet{Xu2008}. They were taken from the Cologne Database for Molecular Spectroscopy (CDMS; \citealt{Muller2001,Muller2005,Endres2016}) and the Jet Propulsion Laboratory catalog \citep{Pickett1998}. $^{(b)}$ Calculated using a collisional temperature of 30~K and collisional rates from \citet{Rabli2010}. The latter were taken from the Leiden Atomic and Molecular Database (LAMDA; \citealt{Schoier2005}).}
\end{tablenotes}
\end{center}
\end{table*}

\section{Observations}
\label{sec:obs}
The Coronet cluster was observed with the SMA \citep{Ho2004} on February 25, 2020, April 28, 30, and May 19, 31, 2021 in the compact configuration, resulting in projected baselines between 17$-$72~m. The region was covered by one pointing centered on IRS7B with coordinates $\alpha_{J2000}=19^\mathrm{h}01^\mathrm{m}56^\mathrm{s}.40$, $\delta_{J2000}=-36^\circ 57\arcmin 27\farcs00$ (Fig.~\ref{fig:sma_continuum}). The SWARM correlator provided a total frequency coverage of 36~GHz, with the lower sideband covering frequencies from 210.3 to 228.3~GHz and upper sideband from 230.3 to 248.3~GHz. The spectral resolution was 0.6~MHz corresponding to 0.7~km~s$^{-1}$.

Data calibration and imaging were performed with the CASA v. 5.7.2 package\footnote{\url{http://casa.nrao.edu/}}\citep{McMullin2007,Bean2022}. The complex gains were calibrated through observations of the quasars 1924-292 and 1957-387 and the bandpass through observations of the bright quasar 3C279. The overall flux calibration was carried out through observations of Vesta and Callisto. Imaging was done using the {\fontfamily{qcr}\selectfont tclean} algorithm and a Briggs weighting with a robust parameter of 0.5. The typical synthesised beam size of the SMA data set was $4\farcs9~\times~2\farcs6$ ($400-700~$AU). The position angle (PA) for CH$_3$OH, our main molecule of interest, at 241.791~GHz was -8.2$^{\circ}$. The PA varies of $\approx 4.5^{\circ}$ across the covered frequencies.

Additional information on the CH$_3$OH large-scale emission towards the Coronet cluster was provided by observations with APEX \citep{Gusten2006} during the nights of October 25 $-$ November 4, 2020. The pointing was the same as reported for the SMA observations. The achieved map size is 150$\arcsec$ $\times$ 150$\arcsec$, fully covering the region observed by the SMA. The frequencies probed by the APEX observations are in the range 235.1~$-$~243~GHz, corresponding to the SMA upper side band observations, with a frequency resolution of 0.061~MHz (0.076~km~s$^{-1}$). The resulting beam-size of the observations is 27.4$\arcsec$ for CH$_3$OH. Data reduction was carried out with the GILDAS package CLASS\footnote{\url{http://www.iram.fr/IRAMFR/GILDAS}}. Subsequently, the reduced APEX and SMA datasets were combined in CASA v. 5.7.2 using the feathering technique, following the procedure presented in Appendix B.1. of \citet{Perotti2020}. The combination of SMA and APEX short-spacing data resulted in mapping the outer regions of protostellar envelopes in the Coronet cluster on scales from $\sim400-10~000$~AU.

\section{Results}
\label{sec:results}
\begin{figure}
\centering
\includegraphics[trim={0 0 0 0}, clip, width=\hsize]{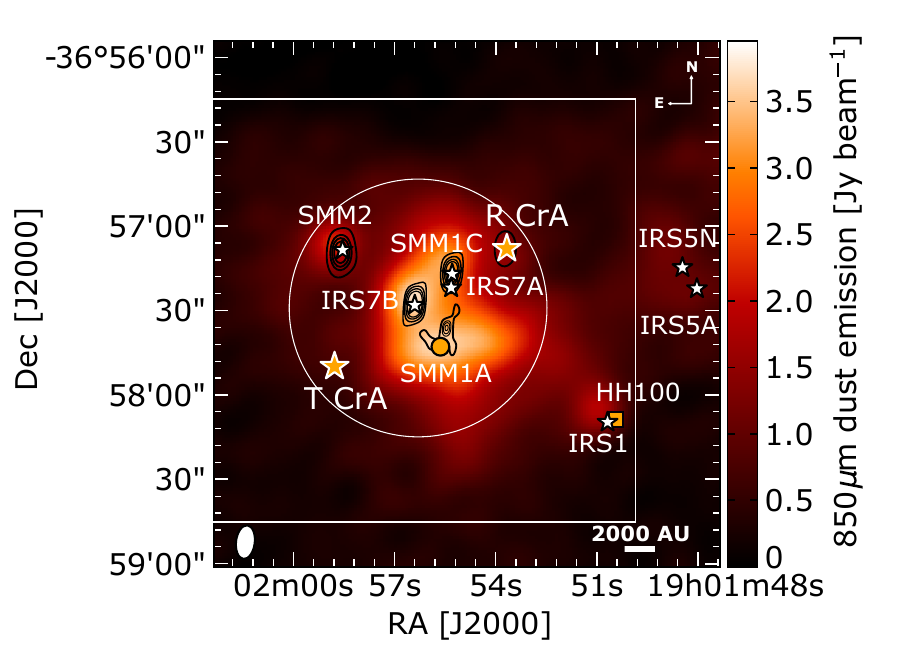}
\caption{SMA continuum at 1.3~mm (contours) overlaid with SCUBA 850~$\mu$m density flux from \citet{Nutter2005}. The contours start at 5$\sigma$ and continue in steps of 5$\sigma$ ($\sigma = 18$~mJy beam$^{-1}$). The empty circle indicates the size of the SMA primary beam, whereas the SMA synthesised beam is shown with a white ellipse in the bottom left corner. The empty rectangle shows the map covered by APEX observations. The stars mark the position of the objects located in the Coronet cluster; refer to Table~\ref{table:samples_sources} for their identification.}
 \label{fig:sma_continuum}
\end{figure}

Figure~\ref{fig:sma_continuum} displays the region targeted by the SMA and APEX observations. The positions of the SMA 1.3~mm continuum peaks coincides with the location of the young stellar objects in the Coronet cluster. The strongest emission is observed towards the two Class~0 sources SMM1C and IRS7B and the Class~I objects SMM2 and IRS7A. Fainter emission north of the pre-stellar core candidate SMM1A \citep{Chen2010} is also seen. Apart from the continuum, line emission of seven species was identified (Table~\ref{table:spectral_data}). These results are presented and discussed in Appendix~\ref{appendixA} (Figures~\ref{mom_zero_part1}$-$\ref{mom_zero_part3}). In the remainder of this section we will uniquely focus on the observed methanol (CH$_3$OH) emission, our molecule of interest.

\begin{figure*}  
\centering
  \includegraphics[trim={0 0 0 0}, clip, width=\hsize]{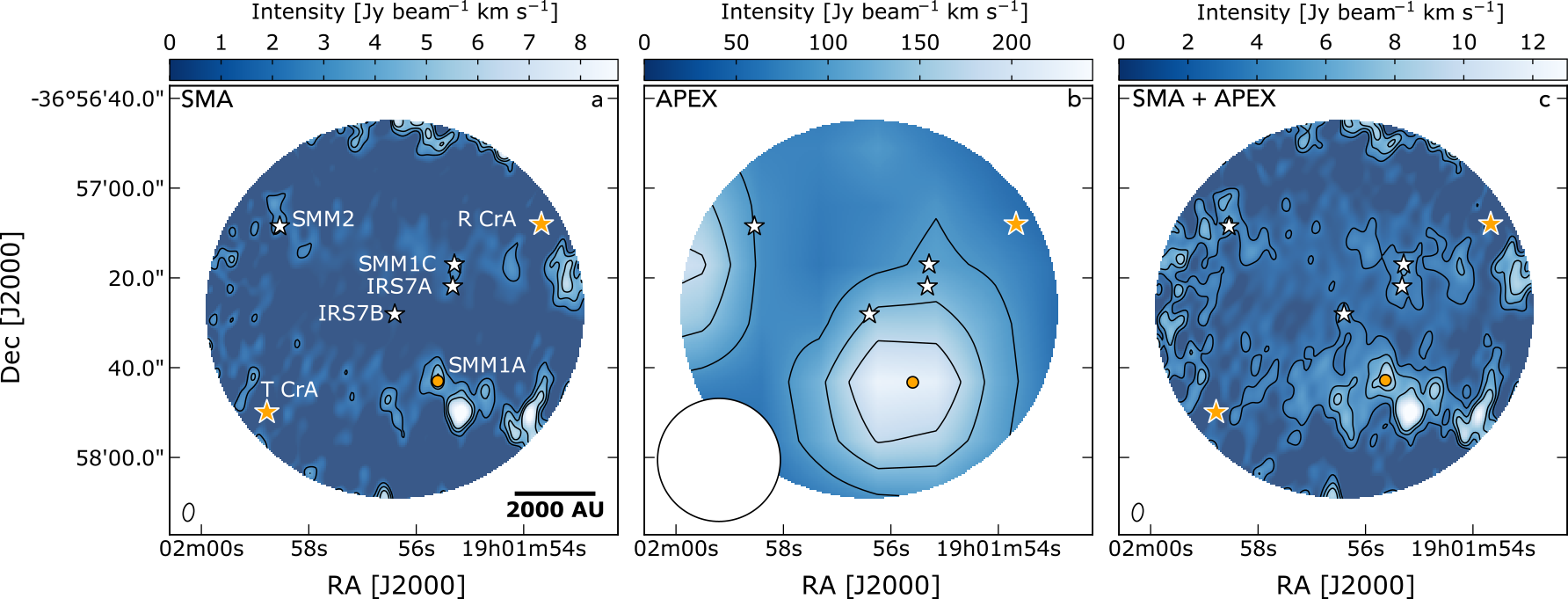}
      \caption{Primary beam corrected integrated intensity maps for CH$_3$OH $J=5_0 - 4_0$~A$^+$ transition ($E_\mathrm{u}=34.8~$K) at 241.791~GHz detected by the SMA (a), by the APEX telescope (b), and in the combined interferometric SMA and single-dish APEX data (c). All lines are integrated between 6 and 12.5~km~s$^{-1}$. Contours start at 5$\sigma$ ($\sigma_\mathrm{SMA}$= 0.16~Jy beam$^{-1}$km~s$^{-1}$, $\sigma_\mathrm{APEX}$= 2~Jy beam$^{-1}$km~s$^{-1}$, $\sigma_\mathrm{SMA+APEX}$= 0.17~Jy beam$^{-1}$km~s$^{-1}$) and follow in steps of 5$\sigma$. The circular field-of-view corresponds to the primary beam of the SMA observations. The synthesised beams are displayed in white in the bottom left corner of each panel.}
         \label{mom0_meth}
\end{figure*}

Amongst the molecular species detected in the Coronet, methanol is the most suited for the main aim of this work: assessing the effect of external irradiation on the gas-phase and solid-state (ice) chemistries in the outer regions of protostellar envelopes. These are the regions probed in the presented SMA and APEX observations, at radii $r>400$~AU from the protostar. Here, gas-phase methanol is not thermally desorbed and it is a tracer of energetic input such as outflows releasing frozen CH$_3$OH to the gas phase (e.g., \citealt{Tychoniec2021}). The latter is due to the fact that the most efficient methanol formation pathway in star-forming regions ($A_\mathrm{V} > 9$ mag) occurs on ice-coated dust grains \citep{Watanabe2002, Fuchs2009,Simons2020,Santos2022}. Consequently, the observed non-thermal CH$_3$OH emission in the outer regions of protostellar envelopes does not trace the bulk of the ices, and instead, it probes a fraction of CH$_3$OH ice sputtered in outflows or photodesorbed (see \citealt{Kristensen2010} for an observational overview of CH$_3$OH non-thermal desorption processes). 

\begin{table*}
\begin{center}
\caption{Total ice and gas column densities ($N$) and fractional abundances ($X$) relative to H$_2$ towards the Coronet cluster members.}
 \label{table:summary_cd}
\renewcommand{\arraystretch}{1.8}
\begin{tabular}{lccccccc} 
\hline \hline
Object & $N^\mathrm{ice}_\mathrm{H_2O}$ $^{(a)}$ & $N^\mathrm{ice}_\mathrm{CH_3OH}$ & $N^\mathrm{gas}_\mathrm{CH_3OH}$ $^{(b)}$ & $N^\mathrm{SCUBA}_\mathrm{H_2}$ $^{(d)}$ & $X^\mathrm{ice}_\mathrm{H_2O}$  & $X^\mathrm{ice}_\mathrm{CH_3OH}$ & $X^\mathrm{gas}_\mathrm{CH_3OH}$ \\   
      &[10$^{18}$cm$^{-2}$] &  [10$^{17}$cm$^{-2}$] & [10$^{15}$cm$^{-2}$]& [10$^{23}$cm$^{-2}$] &[10$^{-5}$] &[10$^{-6}$] & [10$^{-8}$] \\ \hline 
IRS5A         & 3.58 $\pm$ 0.26   & 2.36 $\pm$ 0.57 &   0.05 $\pm$ 0.02$^{(c)}$ & 0.55 $\pm$ 0.06 & 6.51 $\pm$ 0.80   & 4.29 $\pm$ 1.12 & 0.09 $\pm$ 0.04   \\ 
HH100~IRS1    & 2.45 $\pm$ 0.24   & $<$ 2.38        &   0.07 $\pm$ 0.03$^{(c)}$ & 0.67 $\pm$ 0.07 & 3.66 $\pm$ 0.51   & $<$ 3.55        & 0.10 $\pm$ 0.05  \\
SMM1C         &     --            &      --         &   1.95 $\pm$ 0.23           & 1.58 $\pm$ 0.16 &      --             &    --           &  1.23 $\pm$ 0.19 \\
IRS7A         & 10.89 $\pm$ 1.92  & $<$ 4.14        &   1.28 $\pm$ 0.15           & 1.48 $\pm$ 0.15 & 7.36 $\pm$ 1.49     &  $<$ 2.98       &  0.86 $\pm$ 0.13  \\ 
SMM1A         &       --          &      --         &   3.54 $\pm$ 0.43           & 1.90 $\pm$ 0.19 &     --              &    --           &  1.86 $\pm$ 0.29  \\
IRS7B         & 11.01 $\pm$ 1.97  & 7.49 $\pm$ 0.33 &   1.91 $\pm$ 0.23           & 1.92 $\pm$ 0.19 & 5.73 $\pm$ 1.18     & 3.90 $\pm$ 0.43 &  0.99 $\pm$ 0.16   \\
SMM2          &    --             &        --       &   1.64 $\pm$ 0.20           & 1.09 $\pm$ 0.11 &       --            &    --           &  1.50 $\pm$ 0.24     \\
\hline 
\end{tabular}
\end{center}
\footnotesize{\textbf{Notes.} $^{(a)}$\citet{Boogert2008}. $^{(b)}$CH$_3$OH column densities calculated using $T_\mathrm{rot}=30~$K. $^{(c)}$\citet{Lindberg2015}. $^{(d)}$H$_2$ column densities assuming $T_\mathrm{dust}=30~$K. The errors on the H$_2$ column densities are estimated from the 10\% flux calibration uncertainty at 850~$\mu$m. The "--" symbol indicates the sources for which ice column densities and abundances are not available.}
\end{table*}

Figure~\ref{mom0_meth} presents primary beam corrected integrated intensity (moment~0) maps of the CH$_3$OH $J=5_0-4_0$ A$^+$ line for the SMA, APEX, and the combination of interferometric and single-dish data (SMA+APEX). The combined moment~0 maps of the other CH$_3$OH transitions belonging to the $J=5_\mathrm{K}-4_\mathrm{K}$ Q-branch are displayed in Appendix~\ref{appendixA} (Figure~\ref{mom_zero_part2}).
We specifically targeted this branch because its multiple transitions are conveniently observable in a narrow spectral range (0.2~GHz; Table~\ref{table:spectral_data_methanol}) and they are associated to $E_\mathrm{u}$ from 34.8 to 60.7~K, well below the typical CH$_3$OH sublimation temperature as inferred from both observations and laboratory experiments ($70-130$~K; e.g., \citealt{Kristensen2010, Penteado2017}). The latter condition has to be satisfied in order to investigate methanol non-thermal desorption processes. Additionally, observations of CH$_3$OH $J=5_\mathrm{K}-4_\mathrm{K}$ Q-branch make it possible to directly compare our results with previous work on CH$_3$OH gas-to-ice ratios in nearby star-forming regions using the same facilities, observational strategy, and spectral setups \citep{Perotti2020,Perotti2021}. 

By looking at the SMA moment~0 map (Figure~\ref{mom0_meth}; panel a) it is clear that the interferometric observations do not sufficiently recover the spatially extended emission originating from the Coronet cluster. In contrast, they filter out $\approx$90\% of the extended emission detected in the APEX data. Unlike for the other species detected in the SMA data, we here want to perform a quantitative analysis of the CH$_3$OH emission therefore,
we combine the SMA data with the APEX map to avoid underestimating the CH$_3$OH column densities by up to one order of magnitude.

Similarly to the CH$_3$OH $J=4_2-3_1$ E transition (Fig.~\ref{mom_zero_part2}; panel g), the CH$_3$OH $J=5_0-4_0$ A$^+$ emission is tracing the condensations associated to the prestellar core candidate SMM1A. The emission is extended in the APEX map (Fig.~\ref{mom0_meth}; panel b), and confined to two main regions, around SMM1A and to the east of SMM2. The peak intensity is observed also in this case in the vicinity of SMM1A. The same pattern is visible in the SMA~+~APEX moment~0 map (Fig.~\ref{mom0_meth}; panel c), but some CH$_3$OH emission is also present at the SMM2, IRS7A, and IRS7B source positions although approximately a factor of 4 fainter compared to the peak position.

The CH$_3$OH column densities towards the Coronet cluster members were determined from the integrated line intensities of the feathered SMA~+~APEX maps for all five CH$_3$OH transitions belonging to the $J=5_\mathrm{K}-4_\mathrm{K}$ branch (Figure~\ref{mom_zero_part2} and Table~\ref{table:ch3oh_intensities}). The spectra are shown in Figure~\ref{fig:spectra}. For column density calculations, we followed two different treatments: a local thermodynamic equilibrium (LTE) analysis, assuming optically thin emission \citep{Goldsmith1999}, and a non-LTE study employing the RADEX code \citep{vanderTak2007}. For both methods, a temperature of 30~K is assumed \citep{Lindberg2012}. Although the non-LTE analysis indicates that the CH$_3$OH lines are not thermalised (see Figure~\ref{fig:subthermal}) for densities below 10$^7$~cm$^{-3}$, and four out of five transitions are optically thick (see Figure~\ref{fig:tau}), the estimated CH$_3$OH column densities derived with both methods agree with each other (for typical envelope densities $n_\mathrm{H_2}$ between 10$^5$ and 10$^6$~cm$^{-3}$; see Table~\ref{table:comparison_non_LTE}). More details about the LTE and non-LTE analysis are presented in Appendix~\ref{appendixB}. Table~\ref{table:summary_cd} lists the resulting column densities calculated from the LTE analysis and using $T_\mathrm{rot}=30~$K. The uncertainties were calculated from the spectral rms noise and 20$\%$ calibration uncertainty.

\section{Linking methanol ice and gas}
\label{sec:analysis}
In this section, we analyse the gas-ice variations in the Coronet cluster. The H$_2$O and CH$_3$OH ice column densities were determined by \citet{Boogert2008} from \textit{Spitzer} IRS mid-infrared spectra as part of the \textit{Spitzer} Legacy Program "From Molecular Cores to Planet-Forming Disks" (c2d). For the gas-phase counterpart, we made use of CH$_3$OH gas column densities calculated in Sect.~\ref{sec:results} and reported in Table~\ref{table:summary_cd}.

\bigskip
\subsection{H$_2$ column densities}
\label{H2_cd}
Prior to searching for gas-ice correlations, the physical structure of the Coronet cluster is investigated by producing an H$_2$ column density map of the targeted region. This step is required as fractional abundances instead of absolute column densities need to be compared when studying gas-ice variations. This is due to the fact that gas and ice observations, by their very nature, are tracing different spatial scales, and therefore, a comparison between absolute values can lead to misinterpretations (e.g., \citealt{Noble2017}). At the same time it important to note that the H$_2$ column density calculated below provides the total H$_2$ column density - and hence it traces regions where methanol is in the gas phase as well in the ices. As a result, the CH$_3$OH abundances presented in this work have to be interpreted as "average" abundances. 

\begin{figure}[hb!]
\centering
\includegraphics[width=\hsize]{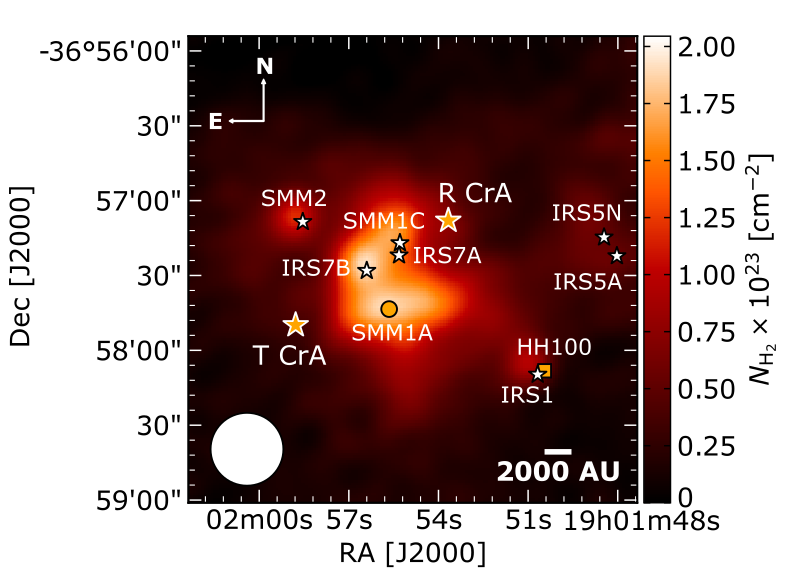}
\caption{H$_2$ column density map of the R~CrA region calculated from SCUBA dust emission maps at 850~$\mu$m by \citet{Nutter2005}. Refer to Fig.~\ref{rgb_rcra} for a guide to the symbols of the Coronet cluster objects.}
 \label{fig:H2_density}
\end{figure}

Figure \ref{fig:H2_density} shows the H$_2$ column density map of the Coronet cluster obtained by making use of SCUBA submillimeter continuum maps at 850~$\mu$m from \citet{Nutter2005}. A detailed description of the formalism adopted to obtain the H$_2$ column density from SCUBA maps is provided in Appendix~C of \citet{Perotti2020}. Briefly, in an optically thin thermal dust emission regime, the strength of the submillimeter radiation depends on the column density ($N$), the dust temperature ($T$) and the opacity ($\kappa_\nu$) \citep{Kauffmann2008}. The inserted value for the opacity per unit dust+gas mass is 0.0182~cm$^2$ g$^{-1}$ ("OH5 dust"; \citealt{Ossenkopf1994}). The adopted value for the dust temperature is 30~K, based on previous observational studies of the Coronet and radiative transfer models by \citet{Lindberg2012}. 

The youngest cluster members are all situated in the densest areas of the targeted region (Fig.~\ref{fig:H2_density}). The apex of the estimated H$_2$ column density lies to the south-east of R~CrA, and in particular at the IRS7B, SMM1A, SMM1C and IRS7A positions ($N_\mathrm{H_2}$ $>$ 1.45$\times$10$^{23}$~cm$^{-2}$). SMM2, IRS1, IRS5N and IRS5A are located in slightly less dense regions (0.5$\times$10$^{23}$~cm$^{-2}$ $<$ $N_\mathrm{H_2}$ $<$ 1.25$\times$10$^{23}$~cm$^{-2}$). The measured H$_2$ column densities and their uncertainties for individual sources are listed in Table~\ref{table:summary_cd}.

\begin{figure*}
\centering
\includegraphics[width=4.5in]{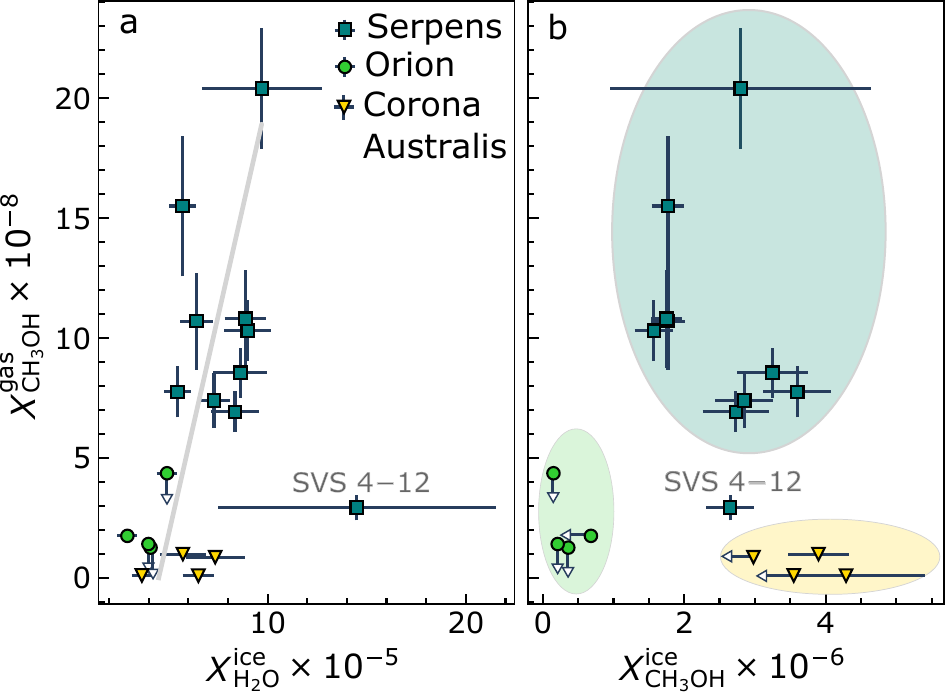}
\caption{Relationship between CH$_3$OH and H$_2$O fractional abundances ($X$) towards three nearby low-mass star-forming regions.
The triangles represent the gas and ice fractional abundances relative to H$_2$ for the Coronet cluster members listed in Table~\ref{table:summary_cd} for which both gas and ice measurements are available. The squares and circles symbolize the values obtained for Serpens SVS~4 and Orion B35A from \citet{Perotti2020} and \citet{Perotti2021}, respectively.
The CH$_3$OH gas abundances are compared to the H$_2$O (panel~a) and CH$_3$OH (panel~b) ice abundances. Upper limits are marked as empty arrows. The grey line in panel~a represents the linear fit to the detections. In panel~b three different groups are identified, excluding Serpens SVS~4$-$12; they correspond to the three star-forming regions and they are marked with coloured ovals.}
 \label{fig:corr_plots1}
\end{figure*}

\subsection{Gas-ice variations}
\label{gas_ice_variations}

The analysis of gas-ice variations in the Coronet is addressed by comparing fractional abundances ($X$) of gas and ice species relative to H$_2$ (Table~\ref{table:summary_cd}). In addition, this section also provides an overview of gas-ice variations in the Coronet with respect to two other nearby low-mass star-forming clusters: Serpens SVS~4 and Orion B35A, which have been observed with the same facilities, angular resolution, and receiver settings adopted for observations of the Coronet to assure a meaningful comparison and reduce observational bias (see \citealt{Perotti2020,Perotti2021}).  
 
Fig.~\ref{fig:corr_plots1} displays CH$_3$OH gas abundances as function of H$_2$O (a) and CH$_3$OH ice (b). A correlation between CH$_3$OH gas and H$_2$O ice for the three star-forming regions is seen in panel a with the Coronet cluster showing gas-ice abundances as low as the Orion B35A cloud. The lower methanol gas abundances in the Coronet and in Orion B35A may be due to reduced ice mantle formation and enhanced photodissociation of methanol molecules upon desorption in these two regions with stronger UV field (see Section~\ref{sec:discussion} for more details). The spread among the H$_2$O abundances for the Coronet cluster and Orion B35A is limited, spanning less than a factor of 2.6 including the uncertainties. The same applies to the CH$_3$OH gas abundances. In contrast, Serpens SVS~4 exhibits the highest values
compared to the other two regions and the largest spread, with CH$_3$OH gas abundances ranging up to $\sim 20 \times 10^{-8}$. One outlier is identified, Serpens SVS~4$-$12, which is characterized by the lowest CH$_3$OH gas and highest H$_2$O ice abundances relative to the other Serpens SVS~4 protostars. This result is not surprising since this is the most extincted object of all targeted sources ($A_\mathrm{V}\sim 95$ mag; \citealt{Pontoppidan2004}) with the lowest S/N (see \citealt{Perotti2020} for a detailed discussion on Serpens SVS~4$-$12).

\begin{figure}
\centering
\includegraphics[width=2.4in]{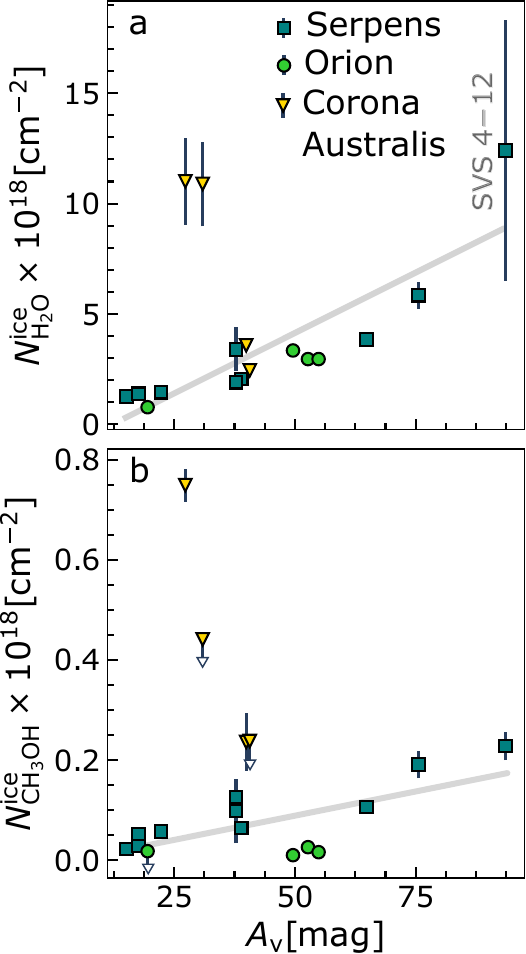}
\caption{Relationship between H$_2$O (a), CH$_3$OH (b) ice column densities ($N$) and visual extinction ($A_\mathrm{V}$) for lines of sights in nearby star-forming clusters: the Coronet in Corona Australis (triangles), Serpens SVS~4 (squares), and Orion B35A (circles). Upper limits are marked as arrows. The gray lines represent linear fits to the detections. We note that the visual extinction values obtained for Corona Australis have to be taken with care as their determination is hampered by the lack of background stars towards the densest region of the Coronet cluster (see Section 4.2). The data for Serpens SVS~4 and Orion B35A were taken from \citet{Perotti2020} and \citet{Perotti2021}, respectively.}
 \label{fig:corr_plots2}
\end{figure}

Fig.~\ref{fig:corr_plots1} b compares CH$_3$OH gas and ice abundances. In contrast to panel a, no clear trend is seen. However, when Serpens SVS~4$-$12 is excluded from the analysis it is possible to identify three different groups corresponding to the three regions: i) Orion B35A with the lowest measurements of both CH$_3$OH in the gas and in the solid state; ii) Serpens SVS~4 with intermediate to high values of CH$_3$OH ice and the highest gas abundances, and finally iii) the Coronet cluster, with some of the highest CH$_3$OH ice but surprisingly low CH$_3$OH gas abundances. The observed behaviour for the Coronet cluster is likely due to CH$_3$OH destruction in the gas phase as a consequence of strong external irradiation \citep{Lindberg2012}.

Figure~\ref{fig:corr_plots2} shows H$_2$O (a) and CH$_3$OH (b) ice column densities ($N$) as function of cloud visual extinction ($A_\mathrm{V}$) for lines of sights in the three nearby star-forming clusters. The cloud extinction estimates were taken from \citet{Pontoppidan2004} for Serpens SVS~4, from \citet{Perotti2021} for Orion B35A, and from \citet{Alves2014} for the Coronet cluster, then converted to visual extinction using the conversion factors (3.6 for $A_J$, 5.55 for $A_H$ and 8.33 for $A_K$) from \citet{Weingartner2001} for RV = 5.5 for the dense ISM. 

The distribution of data points in panels a and b shows a positive trend for Serpens SVS~4 and Orion B35A, indicating that to  higher visual extinctions correspond higher columns of H$_2$O (a) and CH$_3$OH (b). This observation is particularly valid for H$_2$O but it also applies to CH$_3$OH when an evaluation is done per star-forming region.
The opposite behaviour is found for the Coronet cluster members, for both ice species. This negative trend could simply reflect inaccurate extinction values due to the lack of background stars in the densest area of the Corona Australis complex, where the Coronet cluster members are located (see Section~2 of \citet{Alves2014} for more details).

\section{Discussion}
\label{sec:discussion}
In this study, the CH$_3$OH chemical behavior has been investigated with millimetric and infrared facilities. This provides direct observational constraints on the CH$_3$OH gas-to-ice ratio in protostellar envelopes and on its dependency on the physical conditions of star-forming regions. Laboratory experiments predict that CH$_3$OH ice in cold dark clouds is non-thermally desorbed with an efficiency spanning over approximately three orders of magnitude (10$^{-6}-10^{-3}$ molecules/photon; \citealt{Oberg2009b,Bertin2016,Cruz-Diaz2016,Martin-Domenech2016}). This range is expected to increase up to $\sim$10$^{-2}$ molecules/photon \citep{Basalgete2021a,Basalgete2021b}, as X-rays emitted from the central YSO strongly impact the abundances of CH$_3$OH in protostellar envelopes \citep{Notsu2021}. Generally, the majority of non-thermally desorbed CH$_3$OH molecules fragment during the desorption and a fraction of those recombine (e.g., \citealt{Bertin2016}). The CH$_3$OH non-thermal desorption efficiency and by inference the gas-to-ice ratio is highly dependent on the ice structure and composition (pure CH$_3$OH versus CH$_3$OH mixed with CO molecules), as well as on the temperature, photon energy and flux (e.g., \citealt{Oberg2016,Carrascosa2023}). To support laboratory experiments and computations, in this paper we provide observational measurements of the CH$_3$OH gas-to-ice ratio in the outer regions of protostellar envelopes.   

Our targeted regions (Serpens SVS~4, Orion B35A and Corona Australis Coronet) share a number of similarities: e.g., they are low-mass star forming regions, they are clustered and finally, they are affected by the presence of outflows and/or Herbig-Haro objects. However, they show distinct physical conditions for instance, the Serpens SVS~4 cluster is influenced by the presence of the Class~0 binary SMM4 \citep{Pontoppidan2004}, whereas B35A is affected by the nearby high-mass star $\lambda$ Orionis \citep{Reipurth2020} and the Coronet is strongly irradiated by the Herbig Ae/Be star R~CrA \citep{Lindberg2012}. B35A is exposed to an interstellar radiation field \raisebox{2pt}{$\chi_\mathrm{ISRF}$} of 34 \citep{Wolfire1989}\footnote{Note that \citet{Wolfire1989} report a log(G$_0$) of 1.3 which is here converted to \raisebox{2pt}{$\chi_\mathrm{ISRF}$} by applying the formalism described by \citet{Draine_Bertoldi_1996}.} enhanced by the neighbouring $\lambda$ Orionis \citep{Dolan2002}, whereas the radiation field in the Coronet cluster has been estimated to approximately \raisebox{2pt}{$\chi_\mathrm{ISRF}$} $\sim$750 to account for the high fluxes at millimeter wavelengths \citep{Lindberg2012}.

In addition, the molecular clouds in which our targeted regions are located do not share a common formation history. The ongoing low-mass star formation in Serpens Main might have been triggered by cloud-cloud collisions \citep{Duarte2010} or compression from a shock wave from a supernova. However, there is no clear evidence that a nearby supernova ever occurred \citep{Herczeg2019}. In contrast to Serpens, the formation of the ring constituting the $\lambda$ Orionis region is attributed to a supernova explosion that occurred roughly 1$-$6~Myr ago \citep{Dolan1999,Dolan2002,Kounkel2020}. The low-mass star formation in B35A was likely generated by the presence of neighboring massive stars and their stellar winds \citep{Barrado2018}. Finally, star-formation in Corona Australis has been supposedly promoted by a high-velocity cloud impact onto the Galactic plane \citep{Neuhauser2008} or by the expansion of the UpperCenLupus (UCL) superbubble \citep{Mamajek2002}. The fact that the three regions likely did not form in the same way implies different initial physical conditions for the production of CH$_3$OH.

\begin{figure}
\centering
\includegraphics[width=\hsize]{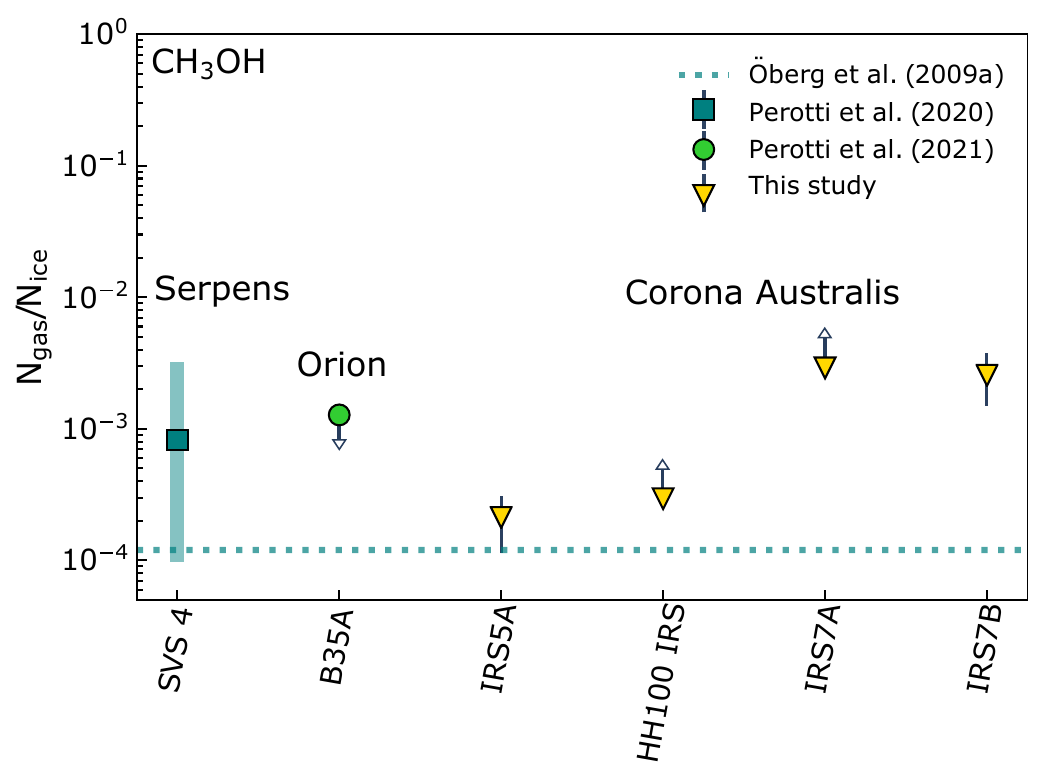}
\caption{CH$_3$OH gas-to-ice ratios ($N_\mathrm{gas}/N_\mathrm{ice}$) towards low-mass protostars in the Coronet cluster in Corona Australis (triangles). The square and the circle symbolize the average ratios for ten sources in Serpens SVS~4 and four sources in Orion B35A, respectively. The green shaded area corresponds to the interval of gas-to-ice ratios observed for the ten sources in the Serpens SVS~4 cluster. The dotted line indicates the average value for four sources from \citet{Oberg2009a}. Upper and lower limits are marked as arrows.}
 \label{fig:gas-to-ice}
\end{figure}

Figure~\ref{fig:gas-to-ice} illustrates the distribution of the CH$_3$OH gas-to-ice ratios ($N_\mathrm{gas}/N_\mathrm{ice}$) towards low-mass protostars located in the different molecular clouds described above. The average CH$_3$OH gas-to-ice ratio (1.2$\times$10$^{-4}$) determined from millimetric (single-dish) and infrared measurements by \citet{Oberg2009a} for four Class 0/I sources in Perseus, Taurus and Serpens is over-plotted. A detailed comparison between the gas-to-ice ratio determined by \citet{Oberg2009a} and the ratios calculated for the sources in the Serpens SVS~4 and the Orion B35A cloud is provided in \citet{Perotti2021}. In this section we focus on the ratios obtained for Corona Australis and their comparison with the values determined for the two other star-forming regions.      

The distribution of CH$_3$OH gas-to-ice ratios towards the Coronet cluster covers approximately one order of magnitude. The uncertainty on the determination of the CH$_3$OH ice column densities for HH100~IRS1 and IRS7A \citep{Boogert2008} results in lower limits for their gas-to-ice ratios, consequently we can not provide solid comparisons for these two sources. In contrast, the values obtained for IRS5A and IRS7B fall in the range of gas-to-ice ratios previously obtained for the protostars located in the Serpens SVS~4 cluster.

The CH$_3$OH gas-to-ice ratios constrained from gas and ice observations of a sample of protostellar envelopes (Figure~\ref{fig:gas-to-ice}) do not point to one value but, instead, to a distribution (1.2$\times$10$^{-4}$ to 3.1$\times$10$^{-3}$) which validates previous observational studies (e.g., \citealt{Oberg2009a,Perotti2020}). This distribution agrees with a more complex scenario for the CH$_3$OH desorption elucidated in the laboratory in which a large fraction of CH$_3$OH molecules do not desorb intact, but instead fragment and eventually recombine during and after the desorption (e.g., \citealt{Bertin2016}) leading to CH$_3$OH gas-to-ice ratios lower than 10$^{-3}$.

A second interesting observation is that the CH$_3$OH gas-to-ice ratios fall in a fairly narrow range; this suggests that the CH$_3$OH chemistry at play in the cold outer regions of protostellar envelopes belonging to different low-mass star-forming regions is relatively independent of variations in the physical conditions. We note that a similar conclusion could not be tested for the innermost envelope regions where thermal desorption dominates and young disks are present; our observations are simply not sensitive to these regions. Our analysis implicitly indicates that the CH$_3$OH chemistry in the outer envelope regions is set to a large degree already during the prestellar stage. A related result came to light when comparing abundances of complex organic molecules observed in distinct star-forming environments (e.g., Galactic disc versus Galactic center, high-mass versus low-mass star-forming region; \citealt{Jorgensen2020,Coletta2020,Yang2021,Nazari2022b}). A larger sample is however necessary to prove this statement, and additional observations, laboratory experiments and modelling efforts are required to elucidate further the impact of the physical evolution of protostars on their chemical signatures and vice versa. 

\section{Conclusions}
\label{sec:conclusions}
We present 1.3~mm SMA and APEX observations towards the Coronet cluster in Corona Australis, a unique astrochemical laboratory to study the effect of external irradiation on the chemical and physical evolution of young protostars. In addition, we determined CH$_3$OH gas-phase abundances and compared those with CH$_3$OH ice abundances from \citet{Boogert2008} to directly constrain the CH$_3$OH gas-to-ice ratio in cold protostellar envelopes. Our key findings are: 
\begin{itemize}

    \item{We find a positive trend between H$_2$O ice and CH$_3$OH gas abundances in Serpens SVS~4, Coronet, and Orion B35A with the latter two regions showing the lowest gas abundances. This result is attributed to reduced ice mantle formation and substantial destruction of methanol gas molecules in these two regions characterized by stronger external UV field.}
  
    \item{The distribution of CH$_3$OH gas-to-ice ratios in the Coronet constrained from millimetric and infrared observations spans over one order of magnitude and reinforces previous observational constraints on the CH$_3$OH gas-to-ice ratio.}

    \item{Similarities are found between CH$_3$OH gas-to-ice ratios determined in different low-mass star forming regions (Serpens SVS~4, Orion B35A and Coronet), characterized by distinct physical conditions and formation histories. This result suggests that the CH$_3$OH chemistry - in the outer regions of low-mass envelopes - is relatively independent of variations in the physical conditions and thus it is set to a large degree during the prestellar stage.} 
\end{itemize}

Multiple avenues for future studies can be pursued. Comparisons of gas and ice abundances of key species are currently limited by the low number of infrared surveys. Current infrared facilities such as the \textit{James Webb} Space Telescope are now routinely probing interstellar ices towards background stars \citep{McClure2023}, pre-stellar cores, proto-stellar envelopes \citep{Yang2022} and near edge-on discs \citep{Sturm2023}, thus significantly increasing the number of studied regions. Therefore, it will be feasible to search for gas-ice trends found in other regions, improve our constraints on the CH$_3$OH gas-to-ice ratios, and ultimately determine ratios for a larger set of major interstellar molecules. This approach will provide important feedback on the interactions between ice and gas material during its journey from the molecular cloud to the disc and on the impact of the physical conditions on the physico-chemical evolution of protostars. In this context, the Corona Australis complex still has much to teach us, hosting one of the youngest population of protostars observed so far. 

\begin{acknowledgements}
We thank Chunhua Qi for translating the SMA data into a format readable in CASA, Henrik Olofsson, and Dirk Petry for guidance in importing the APEX data to CASA and the anonymous reviewer for the careful reading of the manuscript and the useful comments.
This work is based on observations with the Submillimeter Array, Mauna Kea, Hawaii, program codes: 2019B-S014 and 2020B-S058 (PI: Perotti) and with the Atacama Pathfinder Experiment, Llano Chajnantor, Chile, program code: 0105.F-9300 (PI: Perotti). The Submillimeter Array is a joint project between the Smithsonian Astrophysical Observatory and the Academia Sinica Institute of Astronomy and Astrophysics and is funded by the Smithsonian Institution and the Academia Sinica. The Atacama Pathﬁnder EXperiment (APEX) telescope is a collaboration between the Max Planck Institute for Radio Astronomy, the European Southern Observatory, and the Onsala Space Observatory. Swedish observations on APEX are supported through Swedish Research Council grant No 2017-00648. This publication also makes use of data products from the Wide-field Infrared Survey Explorer, which is a joint project of the University of California, Los Angeles, and the Jet Propulsion Laboratory/California Institute of Technology, funded by the National Aeronautics and Space Administration. GP gratefully acknowledges support from the Max Planck Society. The research of MS is supported by NASA under award number 80GSFC21M0002. JKJ acknowledges support from the Independent Research Fund Denmark (grant number 0135-00123B). EAdlV acknowledges financial support provided by FONDECYT grant 3200797. HJF gratefully acknowledges the support of STFC for Astrochemistry at the OU under grant Nos ST/P000584/1 and ST/T005424/1 enabling her participation in this work. SBC was supported  by the NASA Planetary Science Divsion Internal Scientist Funding Program through the Fundamental Laboratory Research work package (FLaRe). 
The authors wish to recognize and acknowledge the very significant cultural role and reverence that the summit of Mauna Kea has always had within the indigenous Hawaiian community. We are most fortunate to have the opportunity to conduct observations from this mountain.
\end{acknowledgements}

\bibliographystyle{aa}
\bibliography{aanda.bbl}

\begin{thebibliography}{111}
\expandafter\ifx\csname natexlab\endcsname\relax\def\natexlab#1{#1}\fi

\bibitem[{{Alves} {et~al.}(2020){Alves}, {Cleeves}, {Girart}, {Zhu}, {Franco},
  {Zurlo}, \& {Caselli}}]{Alves2020}
{Alves}, F.~O., {Cleeves}, L.~I., {Girart}, J.~M., {et~al.} 2020, \apjl, 904,
  L6

\bibitem[{{Alves} {et~al.}(2014){Alves}, {Lombardi}, \& {Lada}}]{Alves2014}
{Alves}, J., {Lombardi}, M., \& {Lada}, C.~J. 2014, \aap, 565, A18

\bibitem[{{Andrews} {et~al.}(2018){Andrews}, {Huang}, {P{\'e}rez}, {Isella},
  {Dullemond}, {Kurtovic}, {Guzm{\'a}n}, {Carpenter}, {Wilner}, {Zhang}, {Zhu},
  {Birnstiel}, {Bai}, {Benisty}, {Hughes}, {{\"O}berg}, \&
  {Ricci}}]{Andrews2018}
{Andrews}, S.~M., {Huang}, J., {P{\'e}rez}, L.~M., {et~al.} 2018, \apjl, 869,
  L41

\bibitem[{{Artur de la Villarmois} {et~al.}(2018){Artur de la Villarmois},
  {Kristensen}, {J{\o}rgensen}, {Bergin}, {Brinch}, {Frimann}, {Harsono},
  {Sakai}, \& {Yamamoto}}]{Artur_2018}
{Artur de la Villarmois}, E., {Kristensen}, L.~E., {J{\o}rgensen}, J.~K.,
  {et~al.} 2018, \aap, 614, A26

\bibitem[{{Balucani} {et~al.}(2015){Balucani}, {Ceccarelli}, \&
  {Taquet}}]{Balucani2015}
{Balucani}, N., {Ceccarelli}, C., \& {Taquet}, V. 2015, \mnras, 449, L16

\bibitem[{{Barrado} {et~al.}(2018){Barrado}, {de Gregorio Monsalvo},
  {Hu{\'e}lamo}, {Morales-Calder{\'o}n}, {Bayo}, {Palau}, {Ruiz},
  {Rivi{\`e}re-Marichalar}, {Bouy}, {Morata}, {Stauffer}, {Eiroa}, \&
  {Noriega-Crespo}}]{Barrado2018}
{Barrado}, D., {de Gregorio Monsalvo}, I., {Hu{\'e}lamo}, N., {et~al.} 2018,
  \aap, 612, A79

\bibitem[{{Basalg{\`e}te} {et~al.}(2021{\natexlab{a}}){Basalg{\`e}te}, {Dupuy},
  {F{\'e}raud}, {Romanzin}, {Philippe}, {Michaut}, {Michoud}, {Amiaud},
  {Lafosse}, {Fillion}, \& {Bertin}}]{Basalgete2021a}
{Basalg{\`e}te}, R., {Dupuy}, R., {F{\'e}raud}, G., {et~al.}
  2021{\natexlab{a}}, \aap, 647, A35

\bibitem[{{Basalg{\`e}te} {et~al.}(2021{\natexlab{b}}){Basalg{\`e}te}, {Dupuy},
  {F{\'e}raud}, {Romanzin}, {Philippe}, {Michaut}, {Michoud}, {Amiaud},
  {Lafosse}, {Fillion}, \& {Bertin}}]{Basalgete2021b}
{Basalg{\`e}te}, R., {Dupuy}, R., {F{\'e}raud}, G., {et~al.}
  2021{\natexlab{b}}, \aap, 647, A36

\bibitem[{Bean {et~al.}(2022)Bean, Bhatnagar, Castro, Meyer, Emonts, Garcia,
  Garwood, Golap, Villalba, Harris, Hayashi, Hoskins, Hsieh, Jagannathan,
  Kawasaki, Keimpema, Kettenis, Lopez, Marvil, Masters, McNichols, Mehringer,
  Miel, Moellenbrock, Montesino, Nakazato, Ott, Petry, Pokorny, Raba, Rau,
  Schiebel, Schweighart, Sekhar, Shimada, Small, Steeb, Sugimoto, Suoranta,
  Tsutsumi, van Bemmel, Verkouter, Wells, Xiong, Szomoru, Griffith,
  Glendenning, \& Kern}]{Bean2022}
Bean, B., Bhatnagar, S., Castro, S., {et~al.} 2022, Publications of the
  Astronomical Society of the Pacific, 134, 114501

\bibitem[{{Bertin} {et~al.}(2016){Bertin}, {Romanzin}, {Doronin}, {Philippe},
  {Jeseck}, {Ligterink}, {Linnartz}, {Michaut}, \& {Fillion}}]{Bertin2016}
{Bertin}, M., {Romanzin}, C., {Doronin}, M., {et~al.} 2016, \apjl, 817, L12

\bibitem[{{Bibo} {et~al.}(1992){Bibo}, {The}, \& {Dawanas}}]{Bibo1992}
{Bibo}, E.~A., {The}, P.~S., \& {Dawanas}, D.~N. 1992, \aap, 260, 293

\bibitem[{{Boogert} {et~al.}(2015){Boogert}, {Gerakines}, \&
  {Whittet}}]{Boogert2015}
{Boogert}, A.~C.~A., {Gerakines}, P.~A., \& {Whittet}, D.~C.~B. 2015, \araa,
  53, 541

\bibitem[{{Boogert} {et~al.}(2008){Boogert}, {Pontoppidan}, {Knez}, {Lahuis},
  {Kessler-Silacci}, {van Dishoeck}, {Blake}, {Augereau}, {Bisschop},
  {Bottinelli}, {Brooke}, {Brown}, {Crapsi}, {Evans}, {Fraser}, {Geers},
  {Huard}, {J{\o}rgensen}, {{\"O}berg}, {Allen}, {Harvey}, {Koerner}, {Mundy},
  {Padgett}, {Sargent}, \& {Stapelfeldt}}]{Boogert2008}
{Boogert}, A.~C.~A., {Pontoppidan}, K.~M., {Knez}, C., {et~al.} 2008, \apj,
  678, 985

\bibitem[{{Bottinelli} {et~al.}(2010){Bottinelli}, {Boogert}, {Bouwman},
  {Beckwith}, {van Dishoeck}, {{\"O}berg}, {Pontoppidan}, {Linnartz}, {Blake},
  {Evans}, \& {Lahuis}}]{Bottinelli2010}
{Bottinelli}, S., {Boogert}, A.~C.~A., {Bouwman}, J., {et~al.} 2010, \apj, 718,
  1100

\bibitem[{{Carrascosa} {et~al.}(2023){Carrascosa}, {Satorre}, {Escribano},
  {Mart{\'\i}n-Dom{\'e}nech}, \& {Mu{\~n}oz Caro}}]{Carrascosa2023}
{Carrascosa}, H., {Satorre}, M.~{\'A}., {Escribano}, B.,
  {Mart{\'\i}n-Dom{\'e}nech}, R., \& {Mu{\~n}oz Caro}, G.~M. 2023, arXiv
  e-prints, arXiv:2308.05980

\bibitem[{{Cazzoletti} {et~al.}(2019){Cazzoletti}, {Manara}, {Baobab Liu}, {van
  Dishoeck}, {Facchini}, {Alcal{\`a}}, {Ansdell}, {Testi}, {Williams},
  {Carrasco-Gonz{\'a}lez}, {Dong}, {Forbrich}, {Fukagawa}, {Galv{\'a}n-Madrid},
  {Hirano}, {Hogerheijde}, {Hasegawa}, {Muto}, {Pinilla}, {Takami}, {Tamura},
  {Tazzari}, \& {Wisniewski}}]{Cazzoletti2019}
{Cazzoletti}, P., {Manara}, C.~F., {Baobab Liu}, H., {et~al.} 2019, \aap, 626,
  A11

\bibitem[{{Chen} \& {Arce}(2010)}]{Chen2010}
{Chen}, X. \& {Arce}, H.~G. 2010, \apjl, 720, L169

\bibitem[{{Chuang} {et~al.}(2016){Chuang}, {Fedoseev}, {Ioppolo}, {van
  Dishoeck}, \& {Linnartz}}]{Chuang2016}
{Chuang}, K.~J., {Fedoseev}, G., {Ioppolo}, S., {van Dishoeck}, E.~F., \&
  {Linnartz}, H. 2016, \mnras, 455, 1702

\bibitem[{{Coletta} {et~al.}(2020){Coletta}, {Fontani}, {Rivilla}, {Mininni},
  {Colzi}, {S{\'a}nchez-Monge}, \& {Beltr{\'a}n}}]{Coletta2020}
{Coletta}, A., {Fontani}, F., {Rivilla}, V.~M., {et~al.} 2020, \aap, 641, A54

\bibitem[{{Cruz-Diaz} {et~al.}(2016){Cruz-Diaz}, {Mart{\'{\i}}n-Dom{\'e}nech},
  {Mu{\~n}oz Caro}, \& {Chen}}]{Cruz-Diaz2016}
{Cruz-Diaz}, G.~A., {Mart{\'{\i}}n-Dom{\'e}nech}, R., {Mu{\~n}oz Caro}, G.~M.,
  \& {Chen}, Y.-J. 2016, \aap, 592, A68

\bibitem[{{Dib} \& {Henning}(2019)}]{Sami2019}
{Dib}, S. \& {Henning}, T. 2019, \aap, 629, A135

\bibitem[{{Dolan} \& {Mathieu}(1999)}]{Dolan1999}
{Dolan}, C.~J. \& {Mathieu}, R.~D. 1999, \aj, 118, 2409

\bibitem[{{Dolan} \& {Mathieu}(2002)}]{Dolan2002}
{Dolan}, C.~J. \& {Mathieu}, R.~D. 2002, \aj, 123, 387

\bibitem[{{Draine} \& {Bertoldi}(1996)}]{Draine_Bertoldi_1996}
{Draine}, B.~T. \& {Bertoldi}, F. 1996, \apj, 468, 269

\bibitem[{{Duarte-Cabral} {et~al.}(2010){Duarte-Cabral}, {Fuller}, {Peretto},
  {Hatchell}, {Ladd}, {Buckle}, {Richer}, \& {Graves}}]{Duarte2010}
{Duarte-Cabral}, A., {Fuller}, G.~A., {Peretto}, N., {et~al.} 2010, \aap, 519,
  A27

\bibitem[{{Endres} {et~al.}(2016){Endres}, {Schlemmer}, {Schilke}, {Stutzki},
  \& {M{\"u}ller}}]{Endres2016}
{Endres}, C.~P., {Schlemmer}, S., {Schilke}, P., {Stutzki}, J., \&
  {M{\"u}ller}, H. S.~P. 2016, Journal of Molecular Spectroscopy, 327, 95

\bibitem[{{Esplin} \& {Luhman}(2022)}]{Esplin2022}
{Esplin}, T.~L. \& {Luhman}, K.~L. 2022, \aj, 163, 64

\bibitem[{{Fedoseev} {et~al.}(2017){Fedoseev}, {Chuang}, {Ioppolo}, {Qasim},
  {van Dishoeck}, \& {Linnartz}}]{Fedoseev2017}
{Fedoseev}, G., {Chuang}, K.~J., {Ioppolo}, S., {et~al.} 2017, \apj, 842, 52

\bibitem[{{Forbrich} \& {Preibisch}(2007)}]{Forbrich2007b}
{Forbrich}, J. \& {Preibisch}, T. 2007, \aap, 475, 959

\bibitem[{{Forbrich} {et~al.}(2007){Forbrich}, {Preibisch}, {Menten},
  {Neuh{\"a}user}, {Walter}, {Tamura}, {Matsunaga}, {Kusakabe}, {Nakajima},
  {Brandeker}, {Fornasier}, {Posselt}, {Tachihara}, \& {Broeg}}]{Forbrich2007a}
{Forbrich}, J., {Preibisch}, T., {Menten}, K.~M., {et~al.} 2007, \aap, 464,
  1003

\bibitem[{{Fuchs} {et~al.}(2009){Fuchs}, {Cuppen}, {Ioppolo}, {Romanzin},
  {Bisschop}, {Andersson}, {van Dishoeck}, \& {Linnartz}}]{Fuchs2009}
{Fuchs}, G.~W., {Cuppen}, H.~M., {Ioppolo}, S., {et~al.} 2009, \aap, 505, 629

\bibitem[{{Galli} {et~al.}(2020){Galli}, {Bouy}, {Olivares}, {Miret-Roig},
  {Sarro}, {Barrado}, {Berihuete}, \& {Brandner}}]{Galli2020}
{Galli}, P.~A.~B., {Bouy}, H., {Olivares}, J., {et~al.} 2020, \aap, 634, A98

\bibitem[{{Garrod} \& {Herbst}(2006)}]{Garrod2006}
{Garrod}, R.~T. \& {Herbst}, E. 2006, \aap, 457, 927

\bibitem[{{Geppert} {et~al.}(2006){Geppert}, {Hamberg}, {Thomas},
  {{\"O}sterdahl}, {Hellberg}, {Zhaunerchyk}, {Ehlerding}, {Millar}, {Roberts},
  {Semaniak}, {Ugglas}, {K{\"a}llberg}, {Simonsson}, {Kaminska}, \&
  {Larsson}}]{Geppert2006}
{Geppert}, W.~D., {Hamberg}, M., {Thomas}, R.~D., {et~al.} 2006, Faraday
  Discussions, 133, 177

\bibitem[{{Goldsmith} \& {Langer}(1999)}]{Goldsmith1999}
{Goldsmith}, P.~F. \& {Langer}, W.~D. 1999, \apj, 517, 209

\bibitem[{{Gray} {et~al.}(2006){Gray}, {Corbally}, {Garrison}, {McFadden},
  {Bubar}, {McGahee}, {O'Donoghue}, \& {Knox}}]{Gray2006}
{Gray}, R.~O., {Corbally}, C.~J., {Garrison}, R.~F., {et~al.} 2006, \aj, 132,
  161

\bibitem[{{Groppi} {et~al.}(2007){Groppi}, {Hunter}, {Blundell}, \&
  {Sandell}}]{Groppi2007}
{Groppi}, C.~E., {Hunter}, T.~R., {Blundell}, R., \& {Sandell}, G. 2007, \apj,
  670, 489

\bibitem[{{G{\"u}sten} {et~al.}(2006){G{\"u}sten}, {Nyman}, {Schilke},
  {Menten}, {Cesarsky}, \& {Booth}}]{Gusten2006}
{G{\"u}sten}, R., {Nyman}, L.~{\AA}., {Schilke}, P., {et~al.} 2006, \aap, 454,
  L13

\bibitem[{{Harsono} {et~al.}(2018){Harsono}, {Bjerkeli}, {van der Wiel},
  {Ramsey}, {Maud}, {Kristensen}, \& {J{\o}rgensen}}]{Harsono2018}
{Harsono}, D., {Bjerkeli}, P., {van der Wiel}, M. H.~D., {et~al.} 2018, Nature
  Astronomy, 2, 646

\bibitem[{{Haworth} {et~al.}(2021){Haworth}, {Kim}, {Winter}, {Hines},
  {Clarke}, {Sellek}, {Ballabio}, \& {Stapelfeldt}}]{Haworth2021}
{Haworth}, T.~J., {Kim}, J.~S., {Winter}, A.~J., {et~al.} 2021, \mnras, 501,
  3502

\bibitem[{{Herbig}(1960)}]{Herbig1960}
{Herbig}, G.~H. 1960, \apjs, 4, 337

\bibitem[{{Herczeg} {et~al.}(2019){Herczeg}, {Kuhn}, {Zhou}, {Hatchell},
  {Manara}, {Johnstone}, {Dunham}, {Bhardwaj}, {Jose}, \& {Yuan}}]{Herczeg2019}
{Herczeg}, G.~J., {Kuhn}, M.~A., {Zhou}, X., {et~al.} 2019, \apj, 878, 111

\bibitem[{{Ho} {et~al.}(2004){Ho}, {Moran}, \& {Lo}}]{Ho2004}
{Ho}, P. T.~P., {Moran}, J.~M., \& {Lo}, K.~Y. 2004, \apjl, 616, L1

\bibitem[{{J{\o}rgensen} {et~al.}(2020){J{\o}rgensen}, {Belloche}, \&
  {Garrod}}]{Jorgensen2020}
{J{\o}rgensen}, J.~K., {Belloche}, A., \& {Garrod}, R.~T. 2020, \araa, 58, 727

\bibitem[{{Kauffmann} {et~al.}(2008){Kauffmann}, {Bertoldi}, {Bourke}, {Evans},
  \& {Lee}}]{Kauffmann2008}
{Kauffmann}, J., {Bertoldi}, F., {Bourke}, T.~L., {Evans}, N.~J., I., \& {Lee},
  C.~W. 2008, \aap, 487, 993

\bibitem[{{Keane} {et~al.}(2001){Keane}, {Tielens}, {Boogert}, {Schutte}, \&
  {Whittet}}]{Keane2001}
{Keane}, J.~V., {Tielens}, A.~G.~G.~M., {Boogert}, A.~C.~A., {Schutte}, W.~A.,
  \& {Whittet}, D.~C.~B. 2001, \aap, 376, 254

\bibitem[{{Keppler} {et~al.}(2018){Keppler}, {Benisty}, {M{\"u}ller},
  {Henning}, {van Boekel}, {Cantalloube}, {Ginski}, {van Holstein}, {Maire},
  {Pohl}, {Samland}, {Avenhaus}, {Baudino}, {Boccaletti}, {de Boer},
  {Bonnefoy}, {Chauvin}, {Desidera}, {Langlois}, {Lazzoni}, {Marleau},
  {Mordasini}, {Pawellek}, {Stolker}, {Vigan}, {Zurlo}, {Birnstiel},
  {Brandner}, {Feldt}, {Flock}, {Girard}, {Gratton}, {Hagelberg}, {Isella},
  {Janson}, {Juhasz}, {Kemmer}, {Kral}, {Lagrange}, {Launhardt}, {Matter},
  {M{\'e}nard}, {Milli}, {Molli{\`e}re}, {Olofsson}, {P{\'e}rez}, {Pinilla},
  {Pinte}, {Quanz}, {Schmidt}, {Udry}, {Wahhaj}, {Williams}, {Buenzli},
  {Cudel}, {Dominik}, {Galicher}, {Kasper}, {Lannier}, {Mesa}, {Mouillet},
  {Peretti}, {Perrot}, {Salter}, {Sissa}, {Wildi}, {Abe}, {Antichi},
  {Augereau}, {Baruffolo}, {Baudoz}, {Bazzon}, {Beuzit}, {Blanchard}, {Brems},
  {Buey}, {De Caprio}, {Carbillet}, {Carle}, {Cascone}, {Cheetham}, {Claudi},
  {Costille}, {Delboulb{\'e}}, {Dohlen}, {Fantinel}, {Feautrier}, {Fusco},
  {Giro}, {Gluck}, {Gry}, {Hubin}, {Hugot}, {Jaquet}, {Le Mignant}, {Llored},
  {Madec}, {Magnard}, {Martinez}, {Maurel}, {Meyer}, {M{\"o}ller-Nilsson},
  {Moulin}, {Mugnier}, {Orign{\'e}}, {Pavlov}, {Perret}, {Petit}, {Pragt},
  {Puget}, {Rabou}, {Ramos}, {Rigal}, {Rochat}, {Roelfsema}, {Rousset}, {Roux},
  {Salasnich}, {Sauvage}, {Sevin}, {Soenke}, {Stadler}, {Suarez}, {Turatto}, \&
  {Weber}}]{Keppler2018}
{Keppler}, M., {Benisty}, M., {M{\"u}ller}, A., {et~al.} 2018, \aap, 617, A44

\bibitem[{{Kounkel}(2020)}]{Kounkel2020}
{Kounkel}, M. 2020, \apj, 902, 122

\bibitem[{{Kristensen} {et~al.}(2010){Kristensen}, {van Dishoeck}, {van
  Kempen}, {Cuppen}, {Brinch}, {J{\o}rgensen}, \&
  {Hogerheijde}}]{Kristensen2010}
{Kristensen}, L.~E., {van Dishoeck}, E.~F., {van Kempen}, T.~A., {et~al.} 2010,
  \aap, 516, A57

\bibitem[{{Kuffmeier} {et~al.}(2020){Kuffmeier}, {Zhao}, \&
  {Caselli}}]{Kuffmeier2020}
{Kuffmeier}, M., {Zhao}, B., \& {Caselli}, P. 2020, \aap, 639, A86

\bibitem[{{Lindberg} \& {J{\o}rgensen}(2012)}]{Lindberg2012}
{Lindberg}, J.~E. \& {J{\o}rgensen}, J.~K. 2012, \aap, 548, A24

\bibitem[{{Lindberg} {et~al.}(2014){Lindberg}, {J{\o}rgensen}, {Brinch},
  {Haugb{\o}lle}, {Bergin}, {Harsono}, {Persson}, {Visser}, \&
  {Yamamoto}}]{Lindberg2014b}
{Lindberg}, J.~E., {J{\o}rgensen}, J.~K., {Brinch}, C., {et~al.} 2014, \aap,
  566, A74

\bibitem[{{Lindberg} {et~al.}(2015){Lindberg}, {J{\o}rgensen}, {Watanabe},
  {Bisschop}, {Sakai}, \& {Yamamoto}}]{Lindberg2015}
{Lindberg}, J.~E., {J{\o}rgensen}, J.~K., {Watanabe}, Y., {et~al.} 2015, \aap,
  584, A28

\bibitem[{{Mamajek} {et~al.}(2002){Mamajek}, {Meyer}, \&
  {Liebert}}]{Mamajek2002}
{Mamajek}, E.~E., {Meyer}, M.~R., \& {Liebert}, J. 2002, \aj, 124, 1670

\bibitem[{{Mart{\'\i}n-Dom{\'e}nech} {et~al.}(2016){Mart{\'\i}n-Dom{\'e}nech},
  {Mu{\~n}oz Caro}, \& {Cruz-D{\'\i}az}}]{Martin-Domenech2016}
{Mart{\'\i}n-Dom{\'e}nech}, R., {Mu{\~n}oz Caro}, G.~M., \& {Cruz-D{\'\i}az},
  G.~A. 2016, \aap, 589, A107

\bibitem[{McClure {et~al.}(2023)McClure, Rocha, Pontoppidan, Crouzet, Chu,
  Dartois, Lamberts, Noble, Pendleton, Perotti, Qasim, Rachid, Smith, Sun,
  Beck, Boogert, Brown, Caselli, Charnley, Cuppen, Dickinson, Drozdovskaya,
  Egami, Erkal, Fraser, Garrod, Harsono, Ioppolo, Jim{\'e}nez-Serra, Jin,
  J{\o}rgensen, Kristensen, Lis, McCoustra, McGuire, Melnick, {\"O}berg,
  Palumbo, Shimonishi, Sturm, van Dishoeck, \& Linnartz}]{McClure2023}
McClure, M.~K., Rocha, W. R.~M., Pontoppidan, K.~M., {et~al.} 2023, Nature
  Astronomy, 7, 431

\bibitem[{{McGuire}(2022)}]{McGuire2022}
{McGuire}, B.~A. 2022, \apjs, 259, 30

\bibitem[{{McMullin} {et~al.}(2007){McMullin}, {Waters}, {Schiebel}, {Young},
  \& {Golap}}]{McMullin2007}
{McMullin}, J.~P., {Waters}, B., {Schiebel}, D., {Young}, W., \& {Golap}, K.
  2007, in Astronomical Society of the Pacific Conference Series, Vol. 376,
  Astronomical Data Analysis Software and Systems XVI, ed. R.~A. {Shaw},
  F.~{Hill}, \& D.~J. {Bell}, 127

\bibitem[{{Mesa} {et~al.}(2019){Mesa}, {Bonnefoy}, {Gratton}, {Van Der Plas},
  {D'Orazi}, {Sissa}, {Zurlo}, {Rigliaco}, {Schmidt}, {Langlois}, {Vigan},
  {Ubeira Gabellini}, {Desidera}, {Antoniucci}, {Barbieri}, {Benisty},
  {Boccaletti}, {Claudi}, {Fedele}, {Gasparri}, {Henning}, {Kasper},
  {Lagrange}, {Lazzoni}, {Lodato}, {Maire}, {Manara}, {Meyer}, {Reggiani},
  {Samland}, {Van den Ancker}, {Chauvin}, {Cheetham}, {Feldt}, {Hugot},
  {Janson}, {Ligi}, {M{\"o}ller-Nilsson}, {Petit}, {Rickman}, {Rigal}, \&
  {Wildi}}]{Mesa2019}
{Mesa}, D., {Bonnefoy}, M., {Gratton}, R., {et~al.} 2019, \aap, 624, A4

\bibitem[{{Miettinen} {et~al.}(2008){Miettinen}, {Kontinen}, {Harju}, \&
  {Higdon}}]{Miettinen2008}
{Miettinen}, O., {Kontinen}, S., {Harju}, J., \& {Higdon}, J.~L. 2008, \aap,
  486, 799

\bibitem[{{M{\"u}ller} {et~al.}(2005){M{\"u}ller}, {Schl{\"o}der}, {Stutzki},
  \& {Winnewisser}}]{Muller2005}
{M{\"u}ller}, H. S.~P., {Schl{\"o}der}, F., {Stutzki}, J., \& {Winnewisser}, G.
  2005, Journal of Molecular Structure, 742, 215

\bibitem[{{M{\"u}ller} {et~al.}(2001){M{\"u}ller}, {Thorwirth}, {Roth}, \&
  {Winnewisser}}]{Muller2001}
{M{\"u}ller}, H.~S.~P., {Thorwirth}, S., {Roth}, D.~A., \& {Winnewisser}, G.
  2001, \aap, 370, L49

\bibitem[{{Nazari} {et~al.}(2022{\natexlab{a}}){Nazari}, {Meijerhof}, {van
  Gelder}, {Ahmadi}, {van Dishoeck}, {Tabone}, {Langeroodi}, {Ligterink},
  {Jaspers}, {Beltr{\'a}n}, {Fuller}, {S{\'a}nchez-Monge}, \&
  {Schilke}}]{Nazari2022b}
{Nazari}, P., {Meijerhof}, J.~D., {van Gelder}, M.~L., {et~al.}
  2022{\natexlab{a}}, \aap, 668, A109

\bibitem[{{Nazari} {et~al.}(2022{\natexlab{b}}){Nazari}, {Tabone}, {Rosotti},
  {van Gelder}, {Meshaka}, \& {van Dishoeck}}]{Nazari2022a}
{Nazari}, P., {Tabone}, B., {Rosotti}, G.~P., {et~al.} 2022{\natexlab{b}},
  \aap, 663, A58

\bibitem[{{Neuh{\"a}user} \& {Forbrich}(2008)}]{Neuhauser2008}
{Neuh{\"a}user}, R. \& {Forbrich}, J. 2008, {The Corona Australis Star Forming
  Region}, Vol.~5 ({Reipurth}, B.), 735

\bibitem[{{Noble} {et~al.}(2017){Noble}, {Fraser}, {Pontoppidan}, \&
  {Craigon}}]{Noble2017}
{Noble}, J.~A., {Fraser}, H.~J., {Pontoppidan}, K.~M., \& {Craigon}, A.~M.
  2017, \mnras, 467, 4753

\bibitem[{{Notsu} {et~al.}(2021){Notsu}, {van Dishoeck}, {Walsh}, {Bosman}, \&
  {Nomura}}]{Notsu2021}
{Notsu}, S., {van Dishoeck}, E.~F., {Walsh}, C., {Bosman}, A.~D., \& {Nomura},
  H. 2021, \aap, 650, A180

\bibitem[{{Nutter} {et~al.}(2005){Nutter}, {Ward-Thompson}, \&
  {Andr{\'e}}}]{Nutter2005}
{Nutter}, D.~J., {Ward-Thompson}, D., \& {Andr{\'e}}, P. 2005, \mnras, 357, 975

\bibitem[{{\"O}berg(2016)}]{Oberg2016}
{\"O}berg, K.~I. 2016, Chemical Reviews, 116, 9631, pMID: 27099922

\bibitem[{{{\"O}berg} {et~al.}(2009{\natexlab{a}}){{\"O}berg}, {Bottinelli}, \&
  {van Dishoeck}}]{Oberg2009a}
{{\"O}berg}, K.~I., {Bottinelli}, S., \& {van Dishoeck}, E.~F.
  2009{\natexlab{a}}, \aap, 494, L13

\bibitem[{{{\"O}berg} {et~al.}(2009{\natexlab{b}}){{\"O}berg}, {Garrod}, {van
  Dishoeck}, \& {Linnartz}}]{Oberg2009b}
{{\"O}berg}, K.~I., {Garrod}, R.~T., {van Dishoeck}, E.~F., \& {Linnartz}, H.
  2009{\natexlab{b}}, \aap, 504, 891

\bibitem[{{Ossenkopf} \& {Henning}(1994)}]{Ossenkopf1994}
{Ossenkopf}, V. \& {Henning}, T. 1994, \aap, 291, 943

\bibitem[{{Penteado} {et~al.}(2017){Penteado}, {Walsh}, \&
  {Cuppen}}]{Penteado2017}
{Penteado}, E.~M., {Walsh}, C., \& {Cuppen}, H.~M. 2017, \apj, 844, 71

\bibitem[{{Perotti} {et~al.}(2021){Perotti}, {J{\o}rgensen}, {Fraser},
  {Suutarinen}, {Kristensen}, {Rocha}, {Bjerkeli}, \&
  {Pontoppidan}}]{Perotti2021}
{Perotti}, G., {J{\o}rgensen}, J.~K., {Fraser}, H.~J., {et~al.} 2021, \aap,
  650, A168

\bibitem[{{Perotti} {et~al.}(2020){Perotti}, {Rocha}, {J{\o}rgensen},
  {Kristensen}, {Fraser}, \& {Pontoppidan}}]{Perotti2020}
{Perotti}, G., {Rocha}, W.~R.~M., {J{\o}rgensen}, J.~K., {et~al.} 2020, \aap,
  643, A48

\bibitem[{{Peterson} {et~al.}(2011){Peterson}, {Caratti o Garatti}, {Bourke},
  {Forbrich}, {Gutermuth}, {J{\o}rgensen}, {Allen}, {Patten}, {Dunham},
  {Harvey}, {Mer{\'\i}n}, {Chapman}, {Cieza}, {Huard}, {Knez}, {Prager}, \&
  {Evans}}]{Peterson2011}
{Peterson}, D.~E., {Caratti o Garatti}, A., {Bourke}, T.~L., {et~al.} 2011,
  \apjs, 194, 43

\bibitem[{{Pickett} {et~al.}(1998){Pickett}, {Poynter}, {Cohen}, {Delitsky},
  {Pearson}, \& {M{\"u}ller}}]{Pickett1998}
{Pickett}, H.~M., {Poynter}, R.~L., {Cohen}, E.~A., {et~al.} 1998, \jqsrt, 60,
  883

\bibitem[{{Pineda} {et~al.}(2023){Pineda}, {Arzoumanian}, {Andre}, {Friesen},
  {Zavagno}, {Clarke}, {Inoue}, {Chen}, {Lee}, {Soler}, \&
  {Kuffmeier}}]{Pineda2023}
{Pineda}, J.~E., {Arzoumanian}, D., {Andre}, P., {et~al.} 2023, in Astronomical
  Society of the Pacific Conference Series, Vol. 534, Astronomical Society of
  the Pacific Conference Series, ed. S.~{Inutsuka}, Y.~{Aikawa}, T.~{Muto},
  K.~{Tomida}, \& M.~{Tamura}, 233

\bibitem[{{Pontoppidan} {et~al.}(2008){Pontoppidan}, {Boogert}, {Fraser}, {van
  Dishoeck}, {Blake}, {Lahuis}, {{\"O}berg}, {Evans}, \&
  {Salyk}}]{Pontoppidan2008}
{Pontoppidan}, K.~M., {Boogert}, A.~C.~A., {Fraser}, H.~J., {et~al.} 2008,
  \apj, 678, 1005

\bibitem[{{Pontoppidan} {et~al.}(2003){Pontoppidan}, {Fraser}, {Dartois},
  {Thi}, {van Dishoeck}, {Boogert}, {d'Hendecourt}, {Tielens}, \&
  {Bisschop}}]{Pontoppidan2003b}
{Pontoppidan}, K.~M., {Fraser}, H.~J., {Dartois}, E., {et~al.} 2003, \aap, 408,
  981

\bibitem[{{Pontoppidan} {et~al.}(2004){Pontoppidan}, {van Dishoeck}, \&
  {Dartois}}]{Pontoppidan2004}
{Pontoppidan}, K.~M., {van Dishoeck}, E.~F., \& {Dartois}, E. 2004, \aap, 426,
  925

\bibitem[{{Qasim} {et~al.}(2018){Qasim}, {Chuang}, {Fedoseev}, {Ioppolo},
  {Boogert}, \& {Linnartz}}]{Qasim2018}
{Qasim}, D., {Chuang}, K.-J., {Fedoseev}, G., {et~al.} 2018, \aap, 612, A83

\bibitem[{{Rabli} \& {Flower}(2010)}]{Rabli2010}
{Rabli}, D. \& {Flower}, D.~R. 2010, \mnras, 406, 95

\bibitem[{{Reipurth} \& {Friberg}(2021)}]{Reipurth2020}
{Reipurth}, B. \& {Friberg}, P. 2021, \mnras, 501, 5938

\bibitem[{{Roberts} \& {Millar}(2000)}]{Roberts2000}
{Roberts}, H. \& {Millar}, T.~J. 2000, \aap, 364, 780

\bibitem[{{Santos} {et~al.}(2022){Santos}, {Chuang}, {Lamberts}, {Fedoseev},
  {Ioppolo}, \& {Linnartz}}]{Santos2022}
{Santos}, J.~C., {Chuang}, K.-J., {Lamberts}, T., {et~al.} 2022, \apjl, 931,
  L33

\bibitem[{{Sch{\"o}ier} {et~al.}(2006){Sch{\"o}ier}, {J{\o}rgensen},
  {Pontoppidan}, \& {Lundgren}}]{Schoier2006}
{Sch{\"o}ier}, F.~L., {J{\o}rgensen}, J.~K., {Pontoppidan}, K.~M., \&
  {Lundgren}, A.~A. 2006, \aap, 454, L67

\bibitem[{{Sch{\"o}ier} {et~al.}(2005){Sch{\"o}ier}, {van der Tak}, {van
  Dishoeck}, \& {Black}}]{Schoier2005}
{Sch{\"o}ier}, F.~L., {van der Tak}, F.~F.~S., {van Dishoeck}, E.~F., \&
  {Black}, J.~H. 2005, \aap, 432, 369

\bibitem[{{Segura-Cox} {et~al.}(2020){Segura-Cox}, {Schmiedeke}, {Pineda},
  {Stephens}, {Fern{\'a}ndez-L{\'o}pez}, {Looney}, {Caselli}, {Li}, {Mundy},
  {Kwon}, \& {Harris}}]{Segura-Cox2020}
{Segura-Cox}, D.~M., {Schmiedeke}, A., {Pineda}, J.~E., {et~al.} 2020, \nat,
  586, 228

\bibitem[{{Shannon} {et~al.}(2014){Shannon}, {Cossou}, {Loison}, {Caubet},
  {Balucani}, {Seakins}, {Wakelam}, \& {Hickson}}]{Shannon2014}
{Shannon}, R.~J., {Cossou}, C., {Loison}, J.-C., {et~al.} 2014, RSC Advances,
  4, 26342

\bibitem[{Sicilia-Aguilar {et~al.}(2011)Sicilia-Aguilar, Henning, Kainulainen,
  \& Roccatagliata}]{Sicilia-Aguilar2011}
Sicilia-Aguilar, A., Henning, T., Kainulainen, J., \& Roccatagliata, V. 2011,
  The Astrophysical Journal, 736, 137

\bibitem[{{Siebenmorgen} \& {Gredel}(1997)}]{Siebenmoregen1997}
{Siebenmorgen}, R. \& {Gredel}, R. 1997, \apj, 485, 203

\bibitem[{{Simons} {et~al.}(2020){Simons}, {Lamberts}, \&
  {Cuppen}}]{Simons2020}
{Simons}, M.~A.~J., {Lamberts}, T., \& {Cuppen}, H.~M. 2020, \aap, 634, A52

\bibitem[{{Sturm} {et~al.}(2023){Sturm}, {McClure}, {Beck}, {Harsono}, B., \&
  {Dartois}}]{Sturm2023}
{Sturm}, J.~A., {McClure}, M.~K., {Beck}, T.~L., {et~al.} 2023, \aap, submitted

\bibitem[{{Taylor} \& {Storey}(1984)}]{Taylor1984}
{Taylor}, K.~N.~R. \& {Storey}, J.~W.~V. 1984, \mnras, 209, 5P

\bibitem[{{Tychoniec} {et~al.}(2020){Tychoniec}, {Manara}, {Rosotti}, {van
  Dishoeck}, {Cridland}, {Hsieh}, {Murillo}, {Segura-Cox}, {van Terwisga}, \&
  {Tobin}}]{Tychoniec2020}
{Tychoniec}, {\L}., {Manara}, C.~F., {Rosotti}, G.~P., {et~al.} 2020, \aap,
  640, A19

\bibitem[{{Tychoniec} {et~al.}(2021){Tychoniec}, {van Dishoeck}, {van't Hoff},
  {van Gelder}, {Tabone}, {Chen}, {Harsono}, {Hull}, {Hogerheijde}, {Murillo},
  \& {Tobin}}]{Tychoniec2021}
{Tychoniec}, {\L}., {van Dishoeck}, E.~F., {van't Hoff}, M. L.~R., {et~al.}
  2021, \aap, 655, A65

\bibitem[{{van der Tak} {et~al.}(2007){van der Tak}, {Black}, {Sch{\"o}ier},
  {Jansen}, \& {van Dishoeck}}]{vanderTak2007}
{van der Tak}, F.~F.~S., {Black}, J.~H., {Sch{\"o}ier}, F.~L., {Jansen}, D.~J.,
  \& {van Dishoeck}, E.~F. 2007, \aap, 468, 627

\bibitem[{{van Dishoeck} \& {Bergin}(2020)}]{vanDishoeck2020}
{van Dishoeck}, E.~F. \& {Bergin}, E.~A. 2020, arXiv e-prints, arXiv:2012.01472

\bibitem[{van Gelder {et~al.}(2022)van Gelder, Nazari, Tabone, Ahmadi, van
  Dishoeck, Beltr{\'{a} }n, Fuller, Sakai, S{\'{a}}nchez-Monge, Schilke, Yang,
  \& Zhang}]{vanGelder2022}
van Gelder, M.~L., Nazari, P., Tabone, B., {et~al.} 2022, \aap, 662, A67

\bibitem[{{van Terwisga} {et~al.}(2020){van Terwisga}, {van Dishoeck}, {Mann},
  {Di Francesco}, {van der Marel}, {Meyer}, {Andrews}, {Carpenter}, {Eisner},
  {Manara}, \& {Williams}}]{vanTerwisga2020}
{van Terwisga}, S.~E., {van Dishoeck}, E.~F., {Mann}, R.~K., {et~al.} 2020,
  \aap, 640, A27

\bibitem[{{Wang} {et~al.}(2004){Wang}, {Mundt}, {Henning}, \&
  {Apai}}]{Wang2004}
{Wang}, H., {Mundt}, R., {Henning}, T., \& {Apai}, D. 2004, \apj, 617, 1191

\bibitem[{{Watanabe} \& {Kouchi}(2002)}]{Watanabe2002}
{Watanabe}, N. \& {Kouchi}, A. 2002, \apjl, 571, L173

\bibitem[{{Watanabe} {et~al.}(2012){Watanabe}, {Sakai}, {Lindberg},
  {J{\o}rgensen}, {Bisschop}, \& {Yamamoto}}]{Watanabe2012}
{Watanabe}, Y., {Sakai}, N., {Lindberg}, J.~E., {et~al.} 2012, \apj, 745, 126

\bibitem[{{Weingartner} \& {Draine}(2001)}]{Weingartner2001}
{Weingartner}, J.~C. \& {Draine}, B.~T. 2001, \apj, 548, 296

\bibitem[{{Winter} {et~al.}(2020){Winter}, {Kruijssen}, {Chevance}, {Keller},
  \& {Longmore}}]{Winter2020}
{Winter}, A.~J., {Kruijssen}, J.~M.~D., {Chevance}, M., {Keller}, B.~W., \&
  {Longmore}, S.~N. 2020, \mnras, 491, 903

\bibitem[{{Wolfire} {et~al.}(1989){Wolfire}, {Hollenbach}, \&
  {Tielens}}]{Wolfire1989}
{Wolfire}, M.~G., {Hollenbach}, D., \& {Tielens}, A.~G.~G.~M. 1989, \apj, 344,
  770

\bibitem[{{Wright} {et~al.}(2010){Wright}, {Eisenhardt}, {Mainzer}, {Ressler},
  {Cutri}, {Jarrett}, {Kirkpatrick}, {Padgett}, {McMillan}, {Skrutskie},
  {Stanford}, {Cohen}, {Walker}, {Mather}, {Leisawitz}, {Gautier}, {McLean},
  {Benford}, {Lonsdale}, {Blain}, {Mendez}, {Irace}, {Duval}, {Liu}, {Royer},
  {Heinrichsen}, {Howard}, {Shannon}, {Kendall}, {Walsh}, {Larsen}, {Cardon},
  {Schick}, {Schwalm}, {Abid}, {Fabinsky}, {Naes}, \& {Tsai}}]{Wright2010}
{Wright}, E.~L., {Eisenhardt}, P. R.~M., {Mainzer}, A.~K., {et~al.} 2010, \aj,
  140, 1868

\bibitem[{Xu {et~al.}(2008)Xu, Fisher, Lees, Shi, Hougen, Pearson, Drouin,
  Blake, \& Braakman}]{Xu2008}
Xu, L.-H., Fisher, J., Lees, R., {et~al.} 2008, J. Mol. Spectrosc., 251, 305

\bibitem[{{Yang} {et~al.}(2022){Yang}, {Green}, {Pontoppidan}, {Bergner},
  {Cleeves}, {Evans}, {Garrod}, {Jin}, {Kim}, {Kim}, {Lee}, {Sakai},
  {Shingledecker}, {Shope}, {Tobin}, \& {van Dishoeck}}]{Yang2022}
{Yang}, Y.-L., {Green}, J.~D., {Pontoppidan}, K.~M., {et~al.} 2022, \apjl, 941,
  L13

\bibitem[{{Yang} {et~al.}(2021){Yang}, {Sakai}, {Zhang}, {Murillo}, {Zhang},
  {Higuchi}, {Zeng}, {L{\'o}pez-Sepulcre}, {Yamamoto}, {Lefloch}, {Bouvier},
  {Ceccarelli}, {Hirota}, {Imai}, {Oya}, {Sakai}, \& {Watanabe}}]{Yang2021}
{Yang}, Y.-L., {Sakai}, N., {Zhang}, Y., {et~al.} 2021, \apj, 910, 20

\end{thebibliography}

\begin{appendix}

\section{Overview of molecular line detections}
\label{appendixA}

This section reports the primary beam corrected integrated intensity (moment~zero) maps of the species detected in the SMA data set (Figures \ref{mom_zero_part1}$-$\ref{mom_zero_part3}). 
For our molecule of interest, CH$_3$OH, we also include in Fig.~\ref{mom_zero_part2} the combined SMA+APEX integrated intensity maps. Please refer to Appendix B.1 of \citet{Perotti2020} for a detailed description of the feathering procedure used to combine interferometric and single-dish data.
 
\begin{table*}[hb!]
\begin{center}
\caption{Identified molecular transitions in the SMA data set.}
\label{table:spectral_data}
\renewcommand{\arraystretch}{1.6}
\begin{tabular}{l l c c c }
\hline \hline
Species & Transition &  Frequency$^{(a)}$ & $A_\mathrm{ul}^{(a)}$  &  $E_\mathrm{u}^{(a)}$  \\ 
       &  & [GHz]   & [s$^{-1}$] & [K]  \\ \hline 
CO-species& &    &   & \\ \hline
$^{12}$CO        & 2~$-$~1         & 230.538   &  6.92~$\times$~10$^{-7}$& 17 \\ 
C$^{18}$O & 2~$-$~1         & 219.560   &  6.01~$\times$~10$^{-7}$& 16 \\ 
$^{13}$CO & 2~$-$~1         & 220.398   &  6.04~$\times~10^{-7}$  & 16  \\
H$_2$CO   & 3$_{0,3}~-~2_{0,2}$   & 218.222   & 2.82~$\times~10^{-4}$   & 21 \\
H$_2$CO   & 3$_{2,2}~-~2_{2,1}$   & 218.476   & 1.58~$\times~10^{-4}$   & 68 \\
H$_2$CO   & 3$_{2,1}~-~2_{2,0}$   & 218.760   & 1.58~$\times~10^{-4}$   & 68 \\
CH$_3$OH  & 4$_2~-~3_1$, E        & 218.440   & 4.68~$\times~10^{-5}$   & 46  \\
CH$_3$OH$^{b}$  & 5$_{+0}~-~4_{+0}$  E & 241.700   &  6.04~$\times~10^{-5}$  & 47.9  \\
CH$_3$OH$^{b}$  & 5$_{-1}~-~4_{-1}$  E & 241.767   &  5.81~$\times~10^{-5}$  & 40.4  \\
CH$_3$OH$^{b}$  & 5$_0~-~4_0$  A$^+$& 241.791   &  6.05~$\times~10^{-5}$  & 34.8  \\
CH$_3$OH$^{b}$  & 5$_{+1}~-~4_{+1}$ E & 241.879   &  5.96~$\times~10^{-5}$ & 55.9   \\
CH$_3$OH$^{b}$  & 5$_{-2}~-~4_{-2}$  E & 241.904   &  5.09~$\times~10^{-5}$  & 60.7    \\ \hline
S-species & &  &  & \\ \hline
SO        & 5$_5~-~4_4$           & 215.220   &    1.20~$\times$~10$^{-4}$                            & 44\\
SO        & 6$_5~-~5_4$          & 219.949   &    1.35~$\times$~10$^{-4}$                            & 35\\ 
SO$_2$    & 4$_{2,2}~-~3_{1,3}$     & 235.151   & 7.69~$\times$~10$^{-5}$ &  19\\ \hline
Carbon-chain species & &  &  & \\ \hline
$c$-C$_3$H$_2$ & 6$_{0,6}~-~5_{1,5}$    & 217.822   &  5.37~$\times$~10$^{-4}$       & 39 \\ 
$c$-C$_3$H$_2$ & 5$_{1,4}~-~4_{2,3}$    & 217.940   &  3.98~$\times$~10$^{-4}$       & 35 \\ 
HC$_3$N &   24~$-$~23           & 218.324  &  8.32~$\times$~10$^{-4}$       & 131 \\
\hline
\hline 
\end{tabular}
\end{center}
\begin{center}
\begin{tablenotes}
\small
\item{\textbf{Notes.}$^{(a)}$ From the Cologne Database for Molecular Spectroscopy (CDMS; \citealt{Muller2001, Muller2005, Endres2016})} and the Jet Propulsion Laboratory catalog \citep{Pickett1998}. The CH$_3$OH transitions belonging to the CH$_3$OH $J_K = 5_K-4_K$ Q-branch are also detected in the APEX data. 
\end{tablenotes}
\end{center}
\end{table*}

The eighteen molecular transitions of seven species identified in the SMA data sets are listed in Table~\ref{table:spectral_data}. The detected species are: carbon monoxide ($^{12}$CO, $^{13}$CO and C$^{18}$O), formaldehyde (H$_2$CO), methanol (CH$_3$OH), sulfur oxide (SO), sulfur dioxide (SO$_2$), cyclopropenylidene (c-C$_3$H$_2$), and cyanoacetylene (HC$_3$N). The line emission is predominantly extended and it does not necessarily coincide with the position of the compact continuum sources in the Coronet cluster (Figures~\ref{mom_zero_part1}- \ref{mom_zero_part3}). In contrast, it is mostly localized to the east of the Herbig Ae/Be star R~CrA and to the south of the young stellar objects IRS7A and IRS7B, at the pre-stellar core SMM1A position. This statement applies especially to H$_2$CO and CH$_3$OH, and is consistent with previous 1.3~mm SMA observations of the region by \citet{Lindberg2012} covering the 216.849$-$218.831~GHz and 226.849$-$228.831~GHz frequencies ranges. Interestingly, no molecular line emission was detected towards SMM1A in the SMA data set presented by \citet{Chen2010}, although the same number of antennas and configuration were used and the SMA correlator covered spectral ranges similar to the ones presented in this work. However, the SMA observations by \citet{Chen2010} were designed to study the dust continuum and they were characterized by a shorter bandwidth and likely lower line sensitivity. Unfortunately, a comparison of the line sensitivity of the two data sets could not be made as this information is not reported in \citet{Chen2010} and a reduced, continuum-subtracted spectrum is not provided either. 

The detected species in the SMA data can be roughly divided into three groups described based on their emission pattern (Table~\ref{table:spectral_data}): (I) CO species, (II) S-species, and (III) carbon-chain species. Overall, the emission reported in the SMA data set is dominated by molecular tracers such as H$_2$CO, CH$_3$OH, SO, SO$_2$. The emission of H$_2$CO and CH$_3$OH is consistent with temperature variations in the field of view of the observations, and particularly with the external irradiation from the variable stars in the region (R~CrA and T~CrA). In contrast, the shock tracers (SO, SO$_2$) probe the numerous outflows and Herbig-Haro objects in the region, which are indicators of a young stellar population in the earliest evolutionary stages.  

\subsection{CO-species}
\begin{figure*}[hb!]
  \centering
  \includegraphics[trim={0 0 0 0}, clip, width=5.8in]{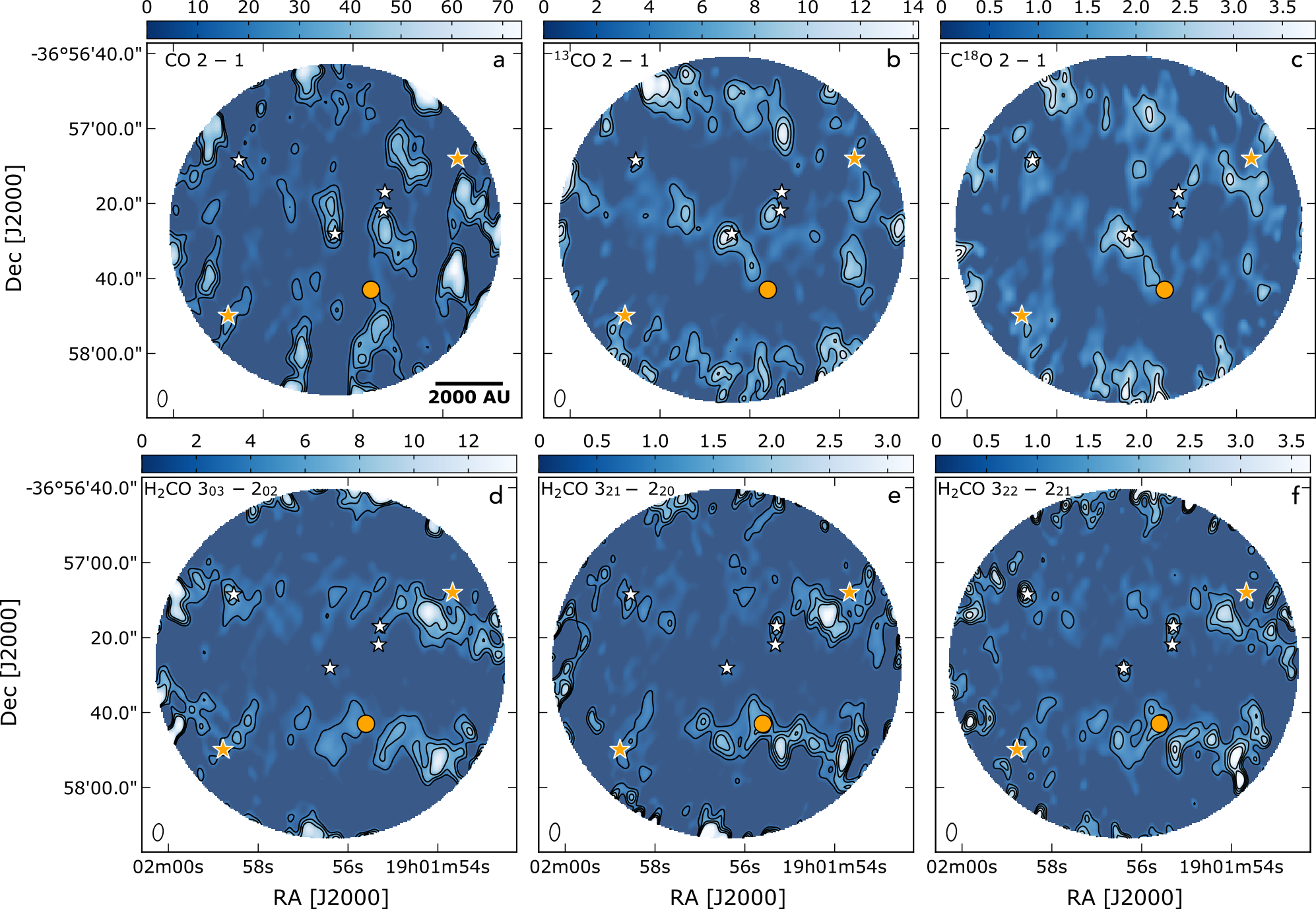}
  \caption{Primary beam corrected integrated intensity maps in Jy beam$^{-1}$km~s$^{-1}$ for the different species detected with the SMA. In each plot contours start at 5$\sigma$ and continue in intervals of 5$\sigma$, except for panel a-c for which contours start at 10$\sigma$ and continue in intervals of 10$\sigma$. Shown are (a) CO $J=2-1$ ($\sigma$= 0.96~Jy beam$^{-1}$km~s$^{-1}$), (b) $^{13}$CO $J=2-1$ ($\sigma$= 0.32~Jy beam$^{-1}$km~s$^{-1}$), (b) C$^{18}$O $J=2-1$ ($\sigma$= 0.13~Jy beam$^{-1}$km~s$^{-1}$),  (d) H$_2$CO 3$_{03}-$2$_{02}$ ($\sigma$= 0.23~mJy beam$^{-1}$km~s$^{-1}$), (e) H$_2$CO 3$_{21}-$2$_{20}$ ($\sigma$= 88~mJy beam$^{-1}$km~s$^{-1}$), (f) H$_2$CO 3$_{22}-$2$_{21}$ ($\sigma$= 0.14~mJy beam$^{-1}$km~s$^{-1}$). The circular field-of-view corresponds to the SMA primary beam. The size of the synthesised beam is shown in the lower left corner of each image. The white and orange stars and the orange circle mark the position of the Coronet cluster members as in Fig.~\ref{rgb_rcra}.}
  \label{mom_zero_part1}
\end{figure*}

Figure~\ref{mom_zero_part1} displays the SMA primary beam corrected integrated intensity maps for $^{12}$CO, $^{13}$CO, C$^{18}$O and H$_2$CO. The distribution of the CO-species emission is extended, especially for $^{12}$CO and its isotopologues.  
The three transitions of H$_2$CO show the same emission pattern (Figure~\ref{mom_zero_part1}, panels d$-$f). The emission is localized in two ridges: the northern ridge between R~CrA and SMM1C, and the southern ridge to the south of IRS7B, in the vicinity of SMM1A. No bright line emission is observed towards the young stellar objects SMM1C, IRS7A and IRS7B. The emission offset from the YSO positions and the elongated shape of the ridges have been previously identified in this region and interpreted as indicators of external heating \citep{Lindberg2012}. The emission localized in the southern ridge traces the three condensations within the pre-stellar core SMM1A (SMM1A-a, SMM1A-b, and SMM1A-c), analysed in dust continuum observations by \citet{Chen2010}.

The strength of the H$_2$CO emission, for the three detected H$_2$CO lines is comparable in both ridges. A similar morphology pattern is observed for CH$_3$OH where the emission peak is located within the SMM1A core. However, neither the CH$_3$OH $J$=4$_2~-~3_1$ line nor the CH$_3$OH transitions belonging to the $J$ = 5$_{K}~-~4_{K}$ branch show a northern ridge (Figures~\ref{mom0_meth} and \ref{mom_zero_part2}).

\subsection{S-species}
The SO and SO$_2$ line emissions appear relatively bright and compact (Figure~\ref{mom_zero_part3}, panels m$-$o). The distribution of the emission is identical for the two SO transitions, and it peaks towards IRS7A and SMM1C. The strength of the emission among the SO transitions is very similar (7$-$8 Jy~beam$^{-1}$~km~s$^{-1}$; Figure~\ref{mom_zero_part3}, panels m$-$n). The SO$_2$ emission is concentrated towards IRS7A and is approximately a factor of 2.5 fainter compared to SO (Figure~\ref{mom_zero_part3}, panels o). In contrast to the majority of the CO-species, the S-species emission have their strongest emission associated with the YSO positions.
SO and SO$_2$ are molecular tracers of energetic inputs in the form of outflows and jets. The luminous SO and SO$_2$ emissions appear to indicate that higher velocity ($\sim5-10$~km~s$^{-1}$) flows of matter are present in the field-of-view of the observations, associated with IRS7A and SMM1C. This result is in agreement with studies of the Herbig-Haro objects in the Coronet attributed to SMM1C \citep{Wang2004,Peterson2011}. 

\subsection{Carbon-chain species}
\begin{figure*}[ht]
  \centering
   \includegraphics[trim={0 0 0 0}, clip, width=5.8in]{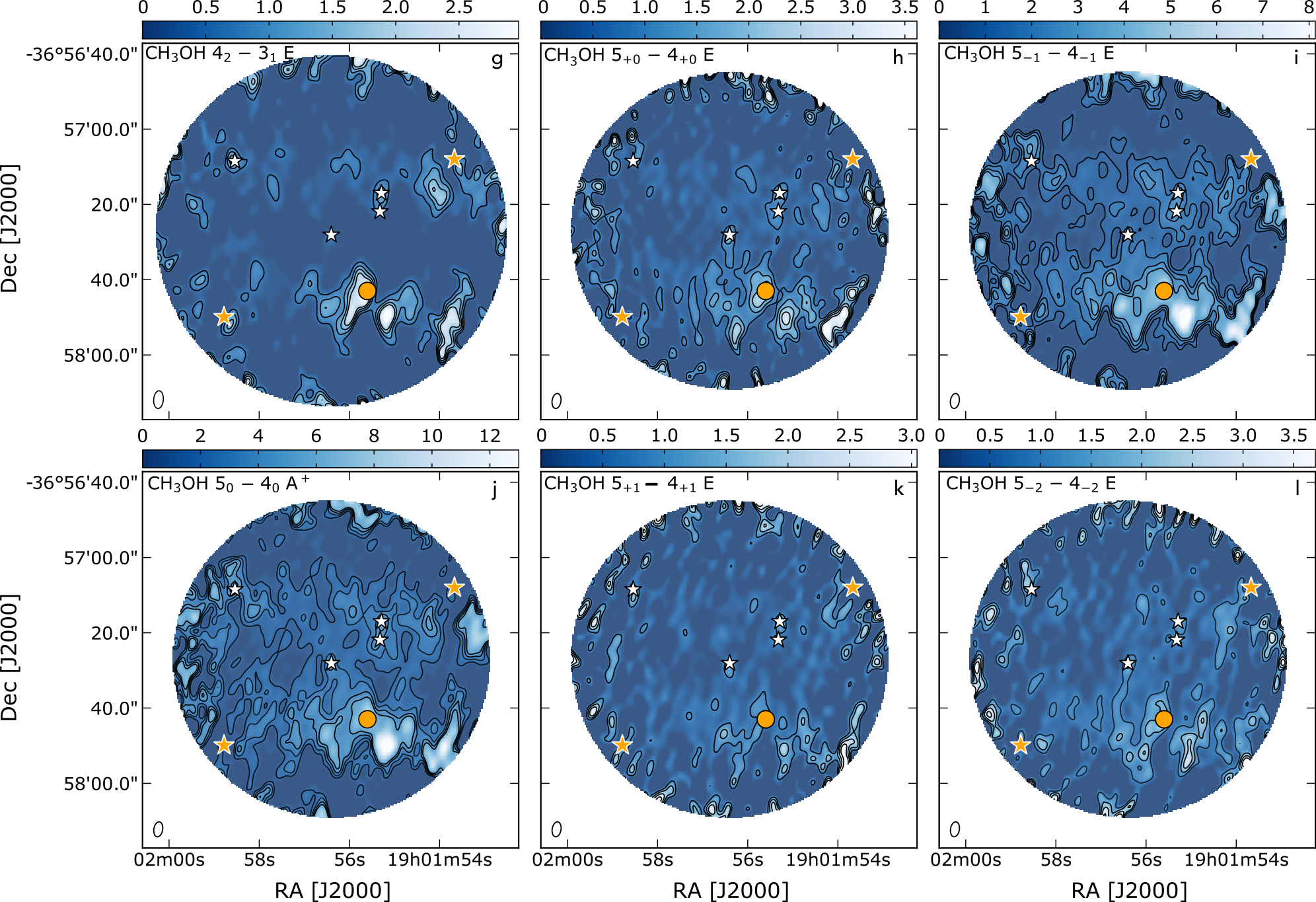}
      \caption{Same as Figure~\ref{mom_zero_part1}. Primary beam corrected integrated intensity maps in Jy beam$^{-1}$km~s$^{-1}$ for: (g) CH$_3$OH 4$_2-$3$_1$ ($\sigma$= 88~mJy beam$^{-1}$km~s$^{-1}$), (h) CH$_3$OH 5$_{+0}-$4$_{+0}$ E ($\sigma$= 0.12~Jy beam$^{-1}$km~s$^{-1}$), (i) CH$_3$OH 5$_{-1}-$4$_{-1}$ E ($\sigma$=0.13~Jy beam$^{-1}$km~s$^{-1}$), (j) CH$_3$OH 5$_{0}-$4$_{0}$ A$^{+}$ ($\sigma$= 0.17~Jy beam$^{-1}$km~s$^{-1}$), (k) CH$_3$OH 5$_{+1}-$4$_{+1}$ E ($\sigma$= 0.13~Jy beam$^{-1}$km~s$^{-1}$), and (l) CH$_3$OH 5$_{-2}-$4$_{-2}$ E ($\sigma$= 0.15~mJy beam$^{-1}$km~s$^{-1}$) detected by the SMA (panel g), and in the combined SMA and APEX data (panels h-l).}
         \label{mom_zero_part2}

  \vspace*{\floatsep}

\includegraphics[trim={0 0 0 0}, clip, width=5.8in]{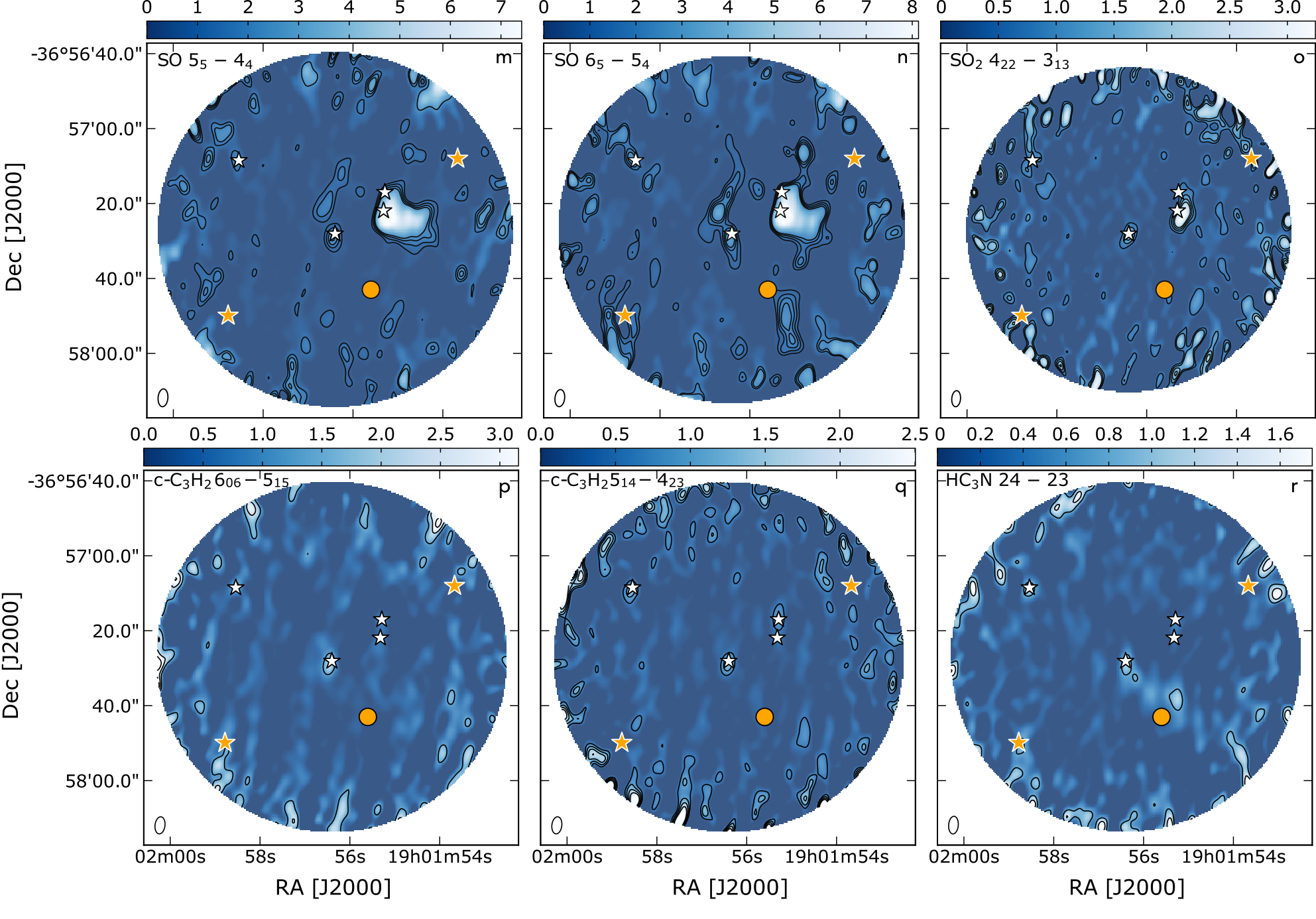}
      \caption{Same as Figure~\ref{mom_zero_part1}. Primary beam corrected integrated intensity maps in Jy beam$^{-1}$km~s$^{-1}$ for : (m) SO $5_5-4_4$ ($\sigma$= 0.12~Jy beam$^{-1}$km~s$^{-1}$), (n) SO $6_5-5_4$ ($\sigma$= 0.13~Jy beam$^{-1}$km~s$^{-1}$), (o) SO$_2$ 4$_{22}-3_{13}$ ($\sigma$= 0.15~Jy beam$^{-1}$km~s$^{-1}$), (p) $c$-C$_3$H$_2$ 6$_{06}-$5$_{15}$ ($\sigma$= 0.13~Jy beam$^{-1}$km~s$^{-1}$), (q) $c$-C$_3$H$_2$ 5$_{14}-$4$_{23}$ ($\sigma$= 0.11~Jy beam$^{-1}$km~s$^{-1}$), and (r) HC$_3$N $J=24-23$ ($\sigma$= 71~mJy beam$^{-1}$km~s$^{-1}$) detected by the SMA.}
         \label{mom_zero_part3}
\end{figure*}

The third group of lines corresponds to carbon-chain species like c-C$_3$H$_2$ and HC$_3$N which show a fairly weak emission (1.6$-$3.0 Jy~beam$^{-1}$~km~s$^{-1}$) compared to the previous two groups (CO- and S-species). Their emission is diffuse and characterized by a fairly low S/N (Figure~\ref{mom_zero_part3}, panels p-r). Similarly to $^{13}$CO and C$^{18}$O, the carbon-chain species have their brightest emission associated with IRS7B. 

\clearpage

\section{Gas-phase methanol}
\label{appendixB}
\subsection{Spectra of methanol}
\begin{figure}[ht!]
\centering
\includegraphics[trim={0 0 0 0}, clip, width=3.5in]{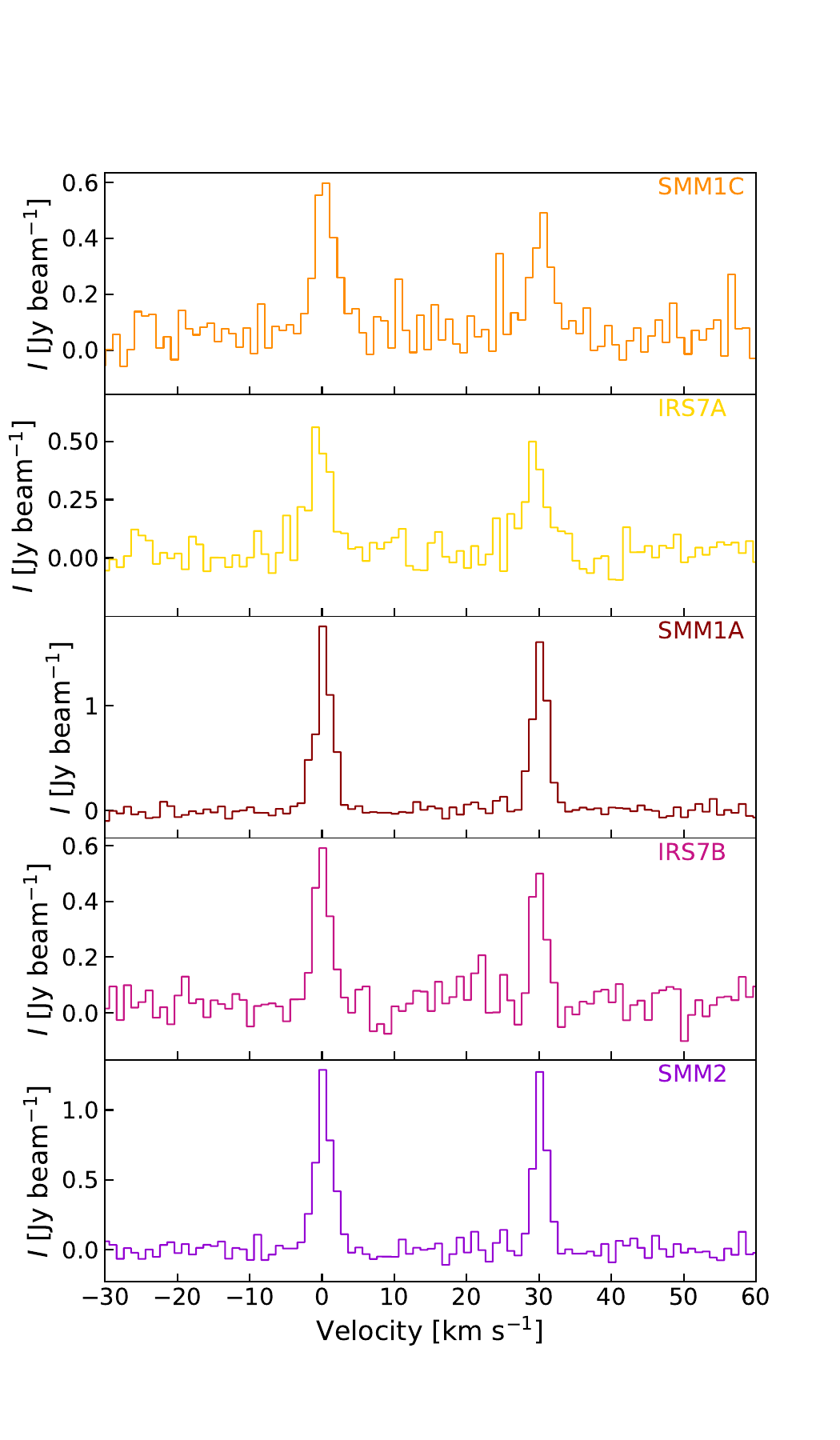}
\caption{CH$_3$OH $J$=5$_0-$4$_0$ A$^{+}$ and CH$_3$OH $J$=5$_{-1}-$4$_{-1}$ E spectra towards the Coronet cluster members detected in the combined SMA+APEX data set.}
\label{fig:spectra}
\end{figure}

Figure~\ref{fig:spectra} shows the spectra for the brightest transitions of the CH$_3$OH $J_K = 5_K-4_K$ Q-branch, $J$=5$_0$ $-$ 4~$_0$~A$^+$ and $J$=5$_{-1}-$4$_{-1}$ E, respectively, extracted from the combined SMA + APEX data set. Table~\ref{table:spectral_data} reports the spectral data for both CH$_3$OH transitions. The CH$_3$OH emission lines have a Gaussian shape and do not show blue or red-shifted components. 

\subsection{Derivation of CH$_3$OH column densities}
The column densities of gas-phase CH$_3$OH for the Coronet cluster members were calculated using the integrated intensities of the combined interferometric (SMA) and single-dish (APEX) data reported in Table~\ref{table:ch3oh_intensities}. The SMA+APEX primary beam corrected integrated intensity (moment~0) maps for the CH$_3$OH $J_K = 5_K-4_K$ Q-branch are displayed in Figure~\ref{mom_zero_part2}. 

\begin{table*}
\begin{center}
\caption{Integrated CH$_3$OH line intensities in units of Jy~beam$^{-1}$~km~s$^{-1}$ towards the Coronet cluster members.}
\label{table:ch3oh_intensities}
\renewcommand{\arraystretch}{1.3}
\begin{tabular}{l c c c c c}
\hline \hline
Source & 5$_{+0}~-~4_{+0}$  E$^+$& 5$_{-1}~-~4_{-1}$  E& 5$_0~-~4_0$  A$^+$& 5$_{+1}~-~4_{+1}$  E& 5$_{-2}~-~4_{-2}$  E \\
\hline
SMM1C  &      1.41$\pm$0.29  &   2.48$\pm$0.50   &   3.86$\pm$0.77   &   1.24$\pm$0.25   &  1.01$\pm$0.21     \\
IRS7A  &      0.79$\pm$0.17  &   2.26$\pm$0.45   &   2.94$\pm$0.59   &   0.61$\pm$0.13   &  0.62$\pm$0.14     \\
SMM1A  &      2.43$\pm$0.49  &   6.44$\pm$1.29   &   7.23$\pm$1.45   &   1.15$\pm$0.24   &  2.57$\pm$0.52      \\
IRS7B  &      1.74$\pm$0.35  &   1.93$\pm$0.39   &  2.97$\pm$0.60   &   0.96$\pm$0.20   &  1.57$\pm$0.32      \\
SMM2  &      0.86$\pm$0.18  &   1.97$\pm$0.40  &   3.79$\pm$0.76   &   0.86$\pm$0.18  &   1.26$\pm$0.26     \\
\hline 
\end{tabular}
\end{center}
\end{table*}

\begin{figure*}
\centering
\includegraphics[trim={0 0 0 0}, clip, width=4in]{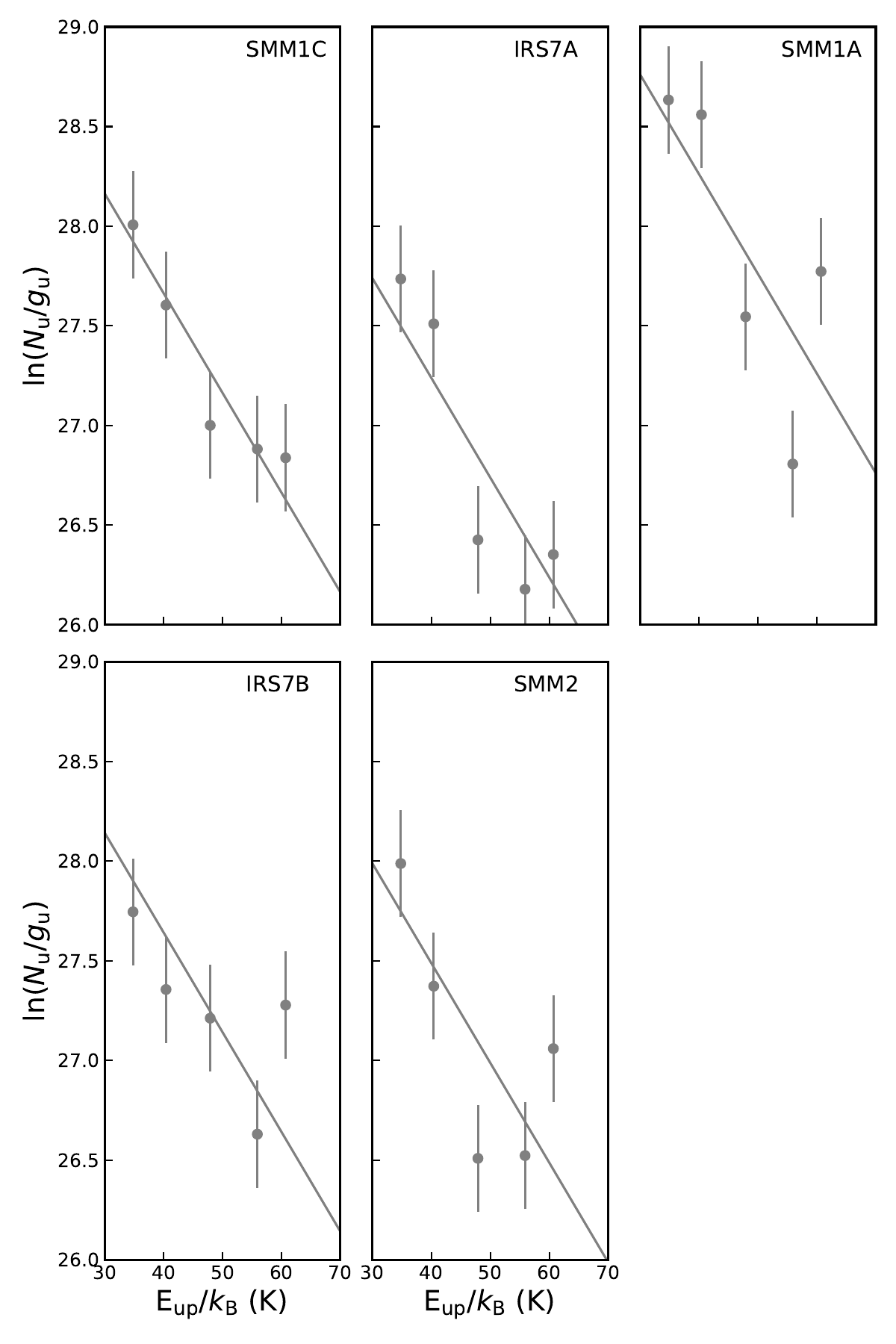}
\caption{Rotational diagrams of CH$_3$OH for the Coronet cluster members. The solid line shows the fixed slope for $T_\mathrm{rot}=30~\mathrm{K}$. Error bars are for 1$\sigma$ uncertainties. The  column densities are reported in Table~\ref{table:summary_cd}. Only transitions belonging to the $J=5_\mathrm{K}-4_\mathrm{K}$ Q-branch have been considered.}
\label{rot_diag}
\end{figure*}

For the column density calculation one of the two adopted procedures assumes LTE conditions, optically thin line emission, homogeneous source filling the telescope beam and a single excitation temperature to describe the level population. The equations used follow the prescription of \citet{Goldsmith1999}, but assuming a fixed rotational temperature equal to 30~K \citep{Lindberg2012}, are reported in Appendix A.4 of \citet{Perotti2021} and thus they will not be repeated here. The values used for the column densities calculation are listed in Table~\ref{table:spectral_data_methanol} and have been compiled from the Cologne Database for Molecular Spectroscopy (CDMS; \citealt{Muller2001,Muller2005,Endres2016}) and the Jet Propulsion Laboratory catalog \citep{Pickett1998}. In particular, the spectroscopic data are from \citet{Xu2008}. The rotational diagrams for the Coronet cluster members are shown in Figure~\ref{rot_diag}. The CH$_3$OH column densities and their uncertainties are reported in Table~\ref{table:summary_cd}. Using a different rotational temperature (e.g., 20 K or 40 K) results in column densities being increased or decreased by approximately a factor of 0.5, respectively. The uncertainties on the gas column densities were calculated based on the spectral rms noise and on the 20\% calibration uncertainty. 

\subsection{Non-LTE Analysis}

\begin{figure*}
\centering
\includegraphics[trim={0 0 0 0}, clip, width=\hsize]{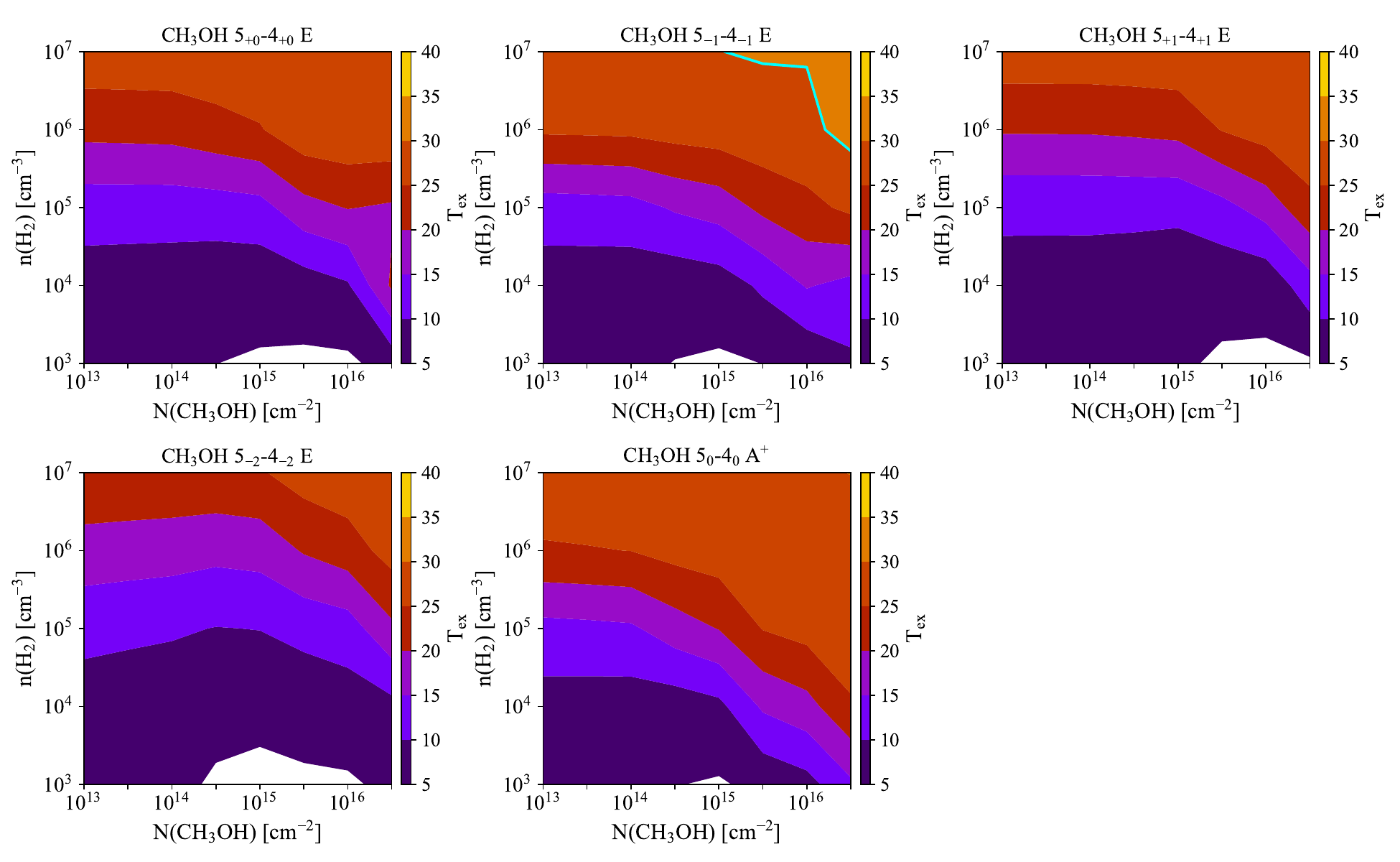}
\caption{RADEX model results for the five CH$_3$OH transitions (Table~\ref{table:spectral_data}) using $T_\mathrm{kin}=30~$K. The cyan line indicates $T_\mathrm{ex}=30~$K.}
\label{fig:subthermal}
\end{figure*}

\begin{figure*}
\centering
\includegraphics[trim={0 0 0 0}, clip, width=\hsize]{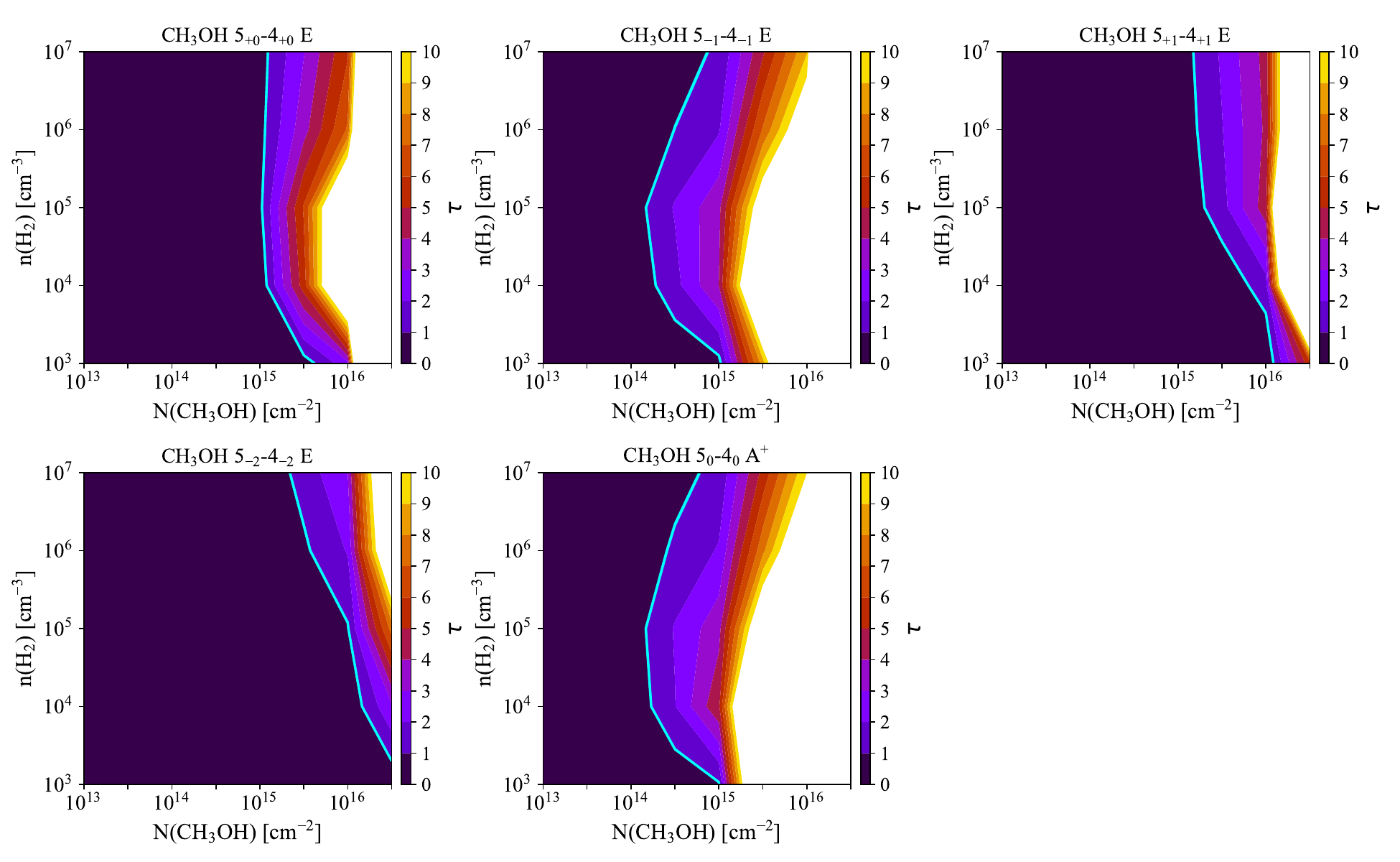}
\caption{Optical depth for the five CH$_3$OH transitions obtained with RADEX, using $T_\mathrm{kin}=30~$K. The cyan line represents $\tau=1$.}
\label{fig:tau}
\end{figure*}

\begin{figure*}
\centering
\begin{subfigure}{.5\textwidth}
  \centering
  \includegraphics[width=1\linewidth]{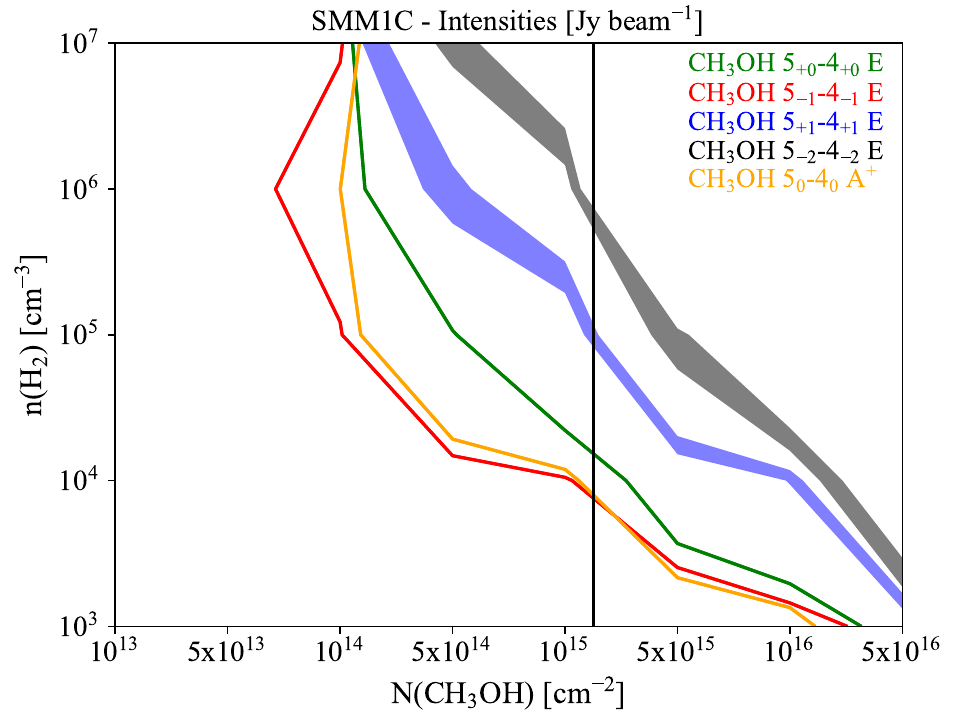}
\end{subfigure}%
\begin{subfigure}{.5\textwidth}
  \centering
  \includegraphics[width=1\linewidth]{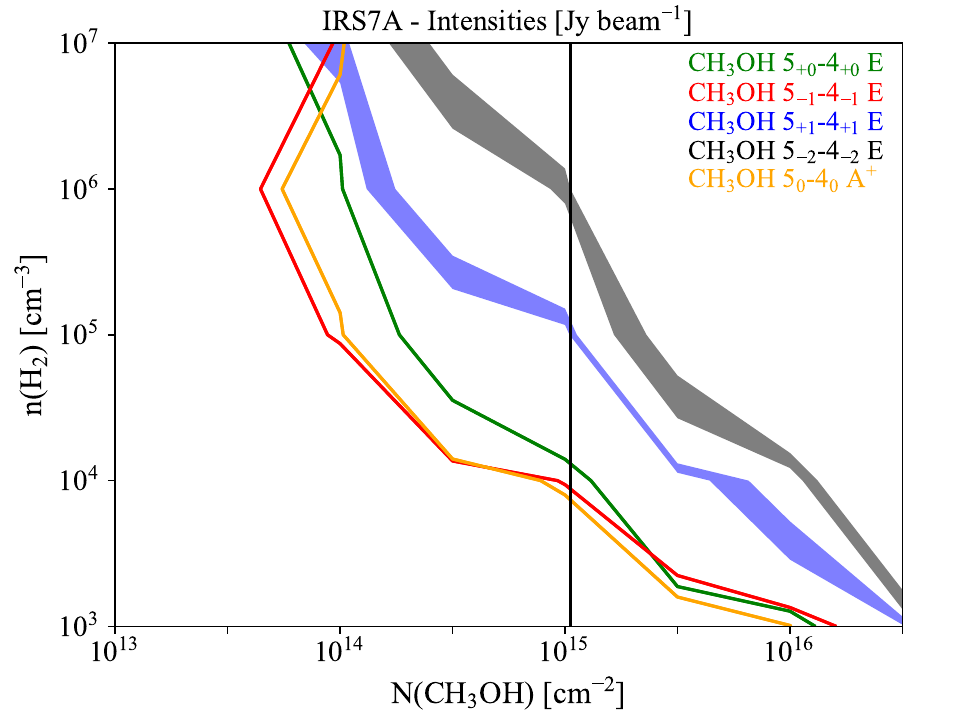}
\end{subfigure}
\caption{RADEX model results for the observed CH$_3$OH lines towards SMM1C (left) and IRS7A (right), using $T_\mathrm{kin}=30~$K. Red, orange and green contours represent lower limits for the integrated line intensities of optically thick lines whereas blue and gray contours represent the marginally optically thin lines and include the errors from Table~\ref{table:ch3oh_intensities}. The vertical line is the CH$_3$OH column density calculated assuming LTE conditions and optically thin emission reported in Table~\ref{table:summary_cd}.}
 \label{fig:appendixB_sources1}
\end{figure*}

\begin{figure*}
\centering
\begin{subfigure}{.5\textwidth}
  \centering
  \includegraphics[width=1\linewidth]{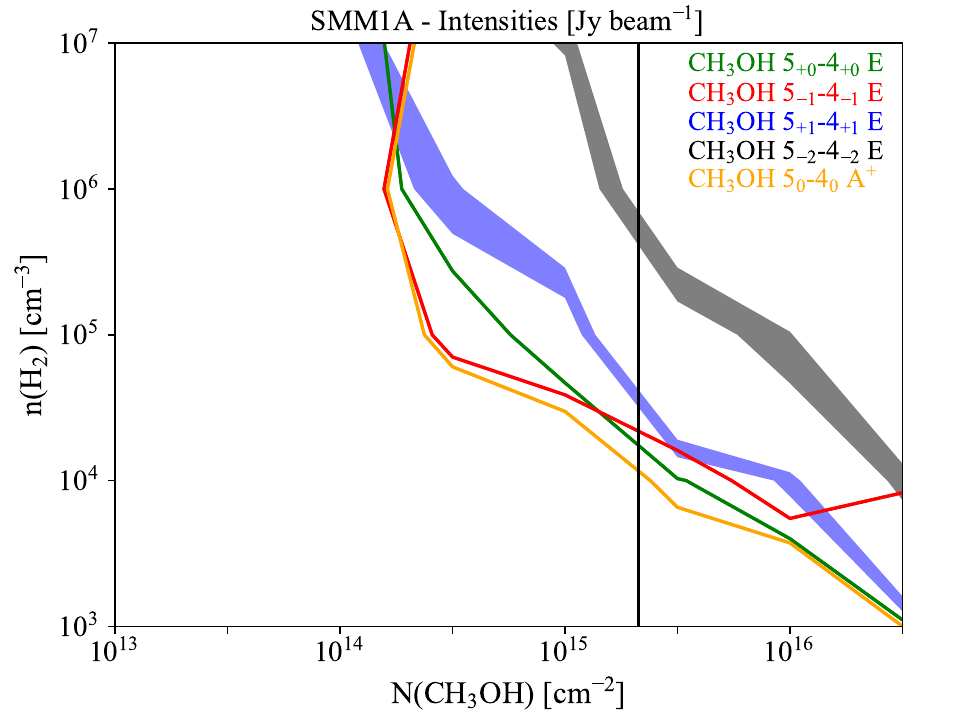}
\end{subfigure}%
\begin{subfigure}{.5\textwidth}
  \centering
  \includegraphics[width=1\linewidth]{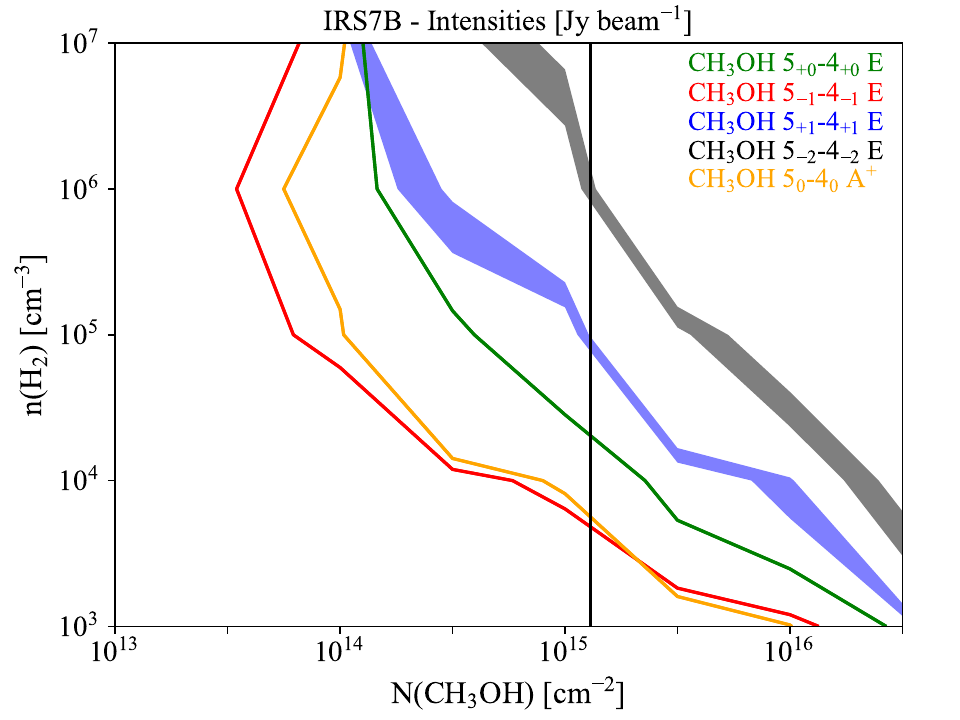}
\end{subfigure}
\caption{Same as Figure~\ref{fig:appendixB_sources1} for SMM1A (left) and IRS7B (right).}
\label{fig:appendixB_sources2}
\end{figure*}

\begin{figure*}
\centering
\begin{subfigure}{.5\textwidth}
  \centering
  \includegraphics[width=1\linewidth]{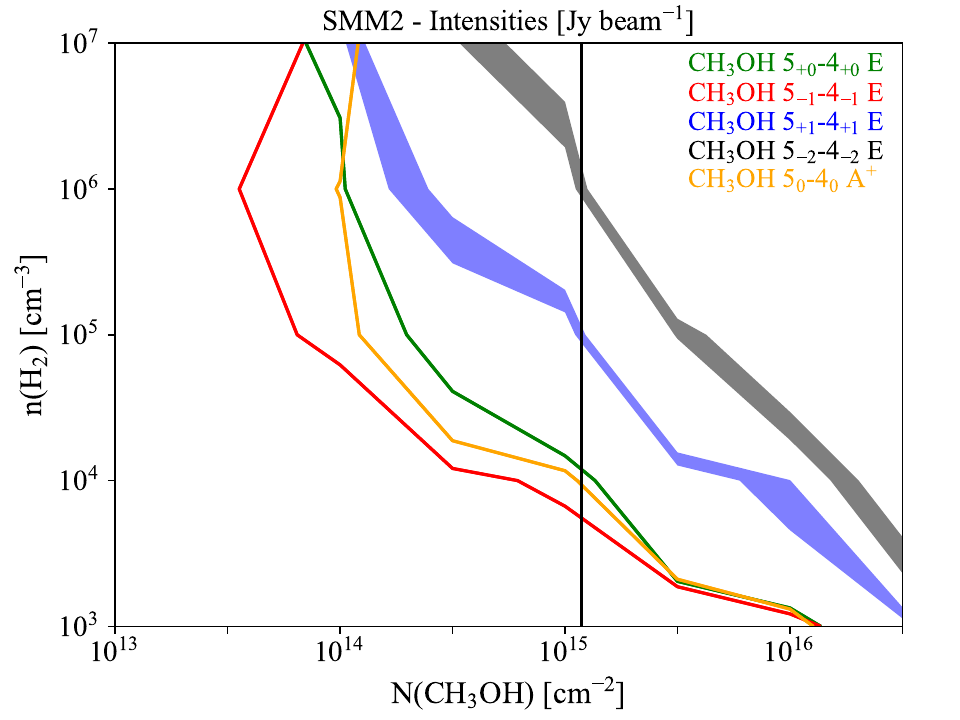}
\end{subfigure}%
\caption{Same as Figure~\ref{fig:appendixB_sources1} for SMM2.}
\label{fig:appendixB_sources3}
\end{figure*}

Further analysis of the CH$_3$OH emission was performed using the non-LTE code RADEX \citep{vanderTak2007}. In particular, all the five  CH$_3$OH transitions were used to compare the observed relative intensities with the RADEX results for a H$_2$ number density ($n_\mathrm{H_2}$) in the range $10^3-10^7~$cm$^{-3}$ and a kinetic temperature equal to 30~K. Figure~\ref{fig:subthermal} shows that all lines are in sub-thermal regime for $n_\mathrm{H_2}$ below $10^7~$cm$^{-3}$, typical for protostellar envelopes.
Figure~\ref{fig:tau} illustrates the optical depth of the five transitions for CH$_3$OH column densities between $10^{13}-10^{16}~$cm$^{-2}$. The three brightest transitions ($J=5_{+0}-4_{+0}$, $J=5_{-1}-4_{-1}$, $J=5_0-4_0$) are optically thick for CH$_3$OH column densities equal to $10^{15}~$cm$^{-2}$, while the $J=5_{+1}-4_{+1}$ transition is marginally optically thick and the weakest transition, $J=5_{-2}-4_{-2}$, can be considered as optically thin. Figures~\ref{fig:appendixB_sources1}$-$~\ref{fig:appendixB_sources3} show a visual comparison between CH$_3$OH column densities calculated using RADEX and obtained from the LTE analysis for each Coronet cluster member. Thin contours represent lower limits for the integrated line intensities of the optically thick lines from Table~\ref{table:ch3oh_intensities}, while thick contours indicate integrated line intensities for the less optically thick lines. The thickness of these contours are given by the errors of the integrated line intensities. A range of CH$_3$OH column densities is estimated for each source from the weakest transition (CH$_3$OH $J=5_{-2}-4_{-2}$) and assuming $n_\mathrm{H_2}$ between 10$^5$ and 10$^6$~cm$^{-3}$ (see Table~\ref{table:comparison_non_LTE}). The vertical black line in Figures~\ref{fig:appendixB_sources1}$-$~\ref{fig:appendixB_sources3} depicts the CH$_3$OH column densities calculated with the LTE analysis listed in Table~\ref{table:summary_cd}. We note that the column densities obtained with both methods are consistent for $n_\mathrm{H_2}$ equal to $10^5-10^6~$cm$^{-3}$ (see Table~\ref{table:comparison_non_LTE}).

\begin{table*}[hb!]
\begin{center}
\caption{Comparison between total CH$_3$OH gas column densities ($N$) calculated using the LTE and the non-LTE analyses towards the Coronet cluster members.}
\label{table:comparison_non_LTE}
\renewcommand{\arraystretch}{1.8}
\begin{tabular}{lcc} 
\hline \hline
Object  & $N^\mathrm{gas, \, LTE}_\mathrm{CH_3OH}$ $^{(a)}$ & $N^\mathrm{gas,\,non-LTE}_\mathrm{CH_3OH}$ $^{(b)}$ \\   
        & [10$^{15}$cm$^{-2}$]             &  [10$^{15}$cm$^{-2}$]                        \\ \hline 
SMM1C   &   1.95 $\pm$ 0.23                &     1.03 $-$ 5.50                            \\
IRS7A   &   1.28 $\pm$ 0.15                &     0.94 $-$ 3.50                            \\ 
SMM1A   &   3.54 $\pm$ 0.43                &     1.50 $-$ 10.0                             \\
IRS7B   &   1.91 $\pm$ 0.23                &     1.07 $-$ 7.30                           \\
SMM2    &   1.64 $\pm$ 0.20                &     1.04 $-$ 6.30                            \\
\hline 
\end{tabular}
\end{center}
\footnotesize{\textbf{Notes.} $^{(a)}$CH$_3$OH column densities calculated with the LTE analysis and fixing $T_\mathrm{rot}=30~$K. $^{(b)}$ CH$_3$OH column densities determined using the non-LTE code RADEX for $T_\mathrm{kin}=30~$K and $n_\mathrm{H_2}=10^5-10^6~$cm$^{-3}$.}
\end{table*}

\end{appendix}
\end{document}